 \documentclass[sigconf]{acmart}

\AtBeginDocument{%
  \providecommand\BibTeX{{%
    \normalfont B\kern-0.5em{\scshape i\kern-0.25em b}\kern-0.8em\TeX}}}

\usepackage{booktabs}
\usepackage{subcaption}
\usepackage{textcomp}
\begin{document}


\title[Who reaps all the Superchats?]{Who Reaps All the Superchats? A Large-Scale Analysis of Income Inequality in Virtual YouTuber Livestreaming}



\author{Ruijing Zhao}
\email{ruijing.zhao@sauder.ubc.ca}
\orcid{0009-0006-2304-2491}
\authornotemark[1]
\affiliation{
    \institution{University of British Columbia}
    \city{Vancouver}
    \country{Canada}
}

\author{Brian Diep}
\authornote{Both authors contributed equally to this research.}
\email{brdiep@mail.ubc.ca}
\orcid{0009-0008-0711-4592}
\affiliation{
    \institution{University of British Columbia}
    \city{Vancouver}
    \country{Canada}
}

\author{Jiaxin Pei}
\email{pedropei@stanford.edu}
\orcid{0000-0002-1849-7962}
\affiliation{
    \institution{Stanford University}
    \city{Palo Alto}
    \country{USA}
}

\author{Dongwook Yoon}
\email{yoon@cs.ubc.ca}
\orcid{0000-0002-7838-8311}
\affiliation{
    \institution{University of British Columbia}
    \city{Vancouver}
    \country{Canada}
}

\author{David Jurgens}
\email{jurgens@umich.edu}
\orcid{0000-0002-2135-9878}
\affiliation{
    \institution{University of Michigan}
    \city{Ann Arbor}
    \country{USA}
}

\author{Jian Zhu}
\email{jian.zhu@ubc.ca}
\orcid{0000-0002-7849-1060}
\affiliation{
    \institution{University of British Columbia}
    \city{Vancouver}
    \country{Canada}
}



\begin{abstract}
The explosive growth of Virtual YouTubers (VTubers)---streamers who perform behind virtual anime avatars---has created a unique digital economy with profound implications for content creators, platforms, and viewers. Understanding the economic landscape of VTubers is crucial for designing equitable platforms, supporting content creator livelihoods, and fostering sustainable digital communities. To this end, we conducted a large-scale study of over 1 million hours of publicly available streaming records from 1,923 VTubers on YouTube, covering tens of millions of dollars in actual profits. 
Our analysis reveals stark inequality within the VTuber community and characterizes the sources of income for VTubers from multiple perspectives. Furthermore, we also found that the VTuber community is increasingly monopolized by two agencies, driving the financial disparity. 
This research illuminates the financial dynamics of VTuber communities, informing the design of equitable platforms and sustainable support systems for digital content creators.

\end{abstract}





\maketitle

\begin{figure*}
    \centering
    \includegraphics[width=1\linewidth]{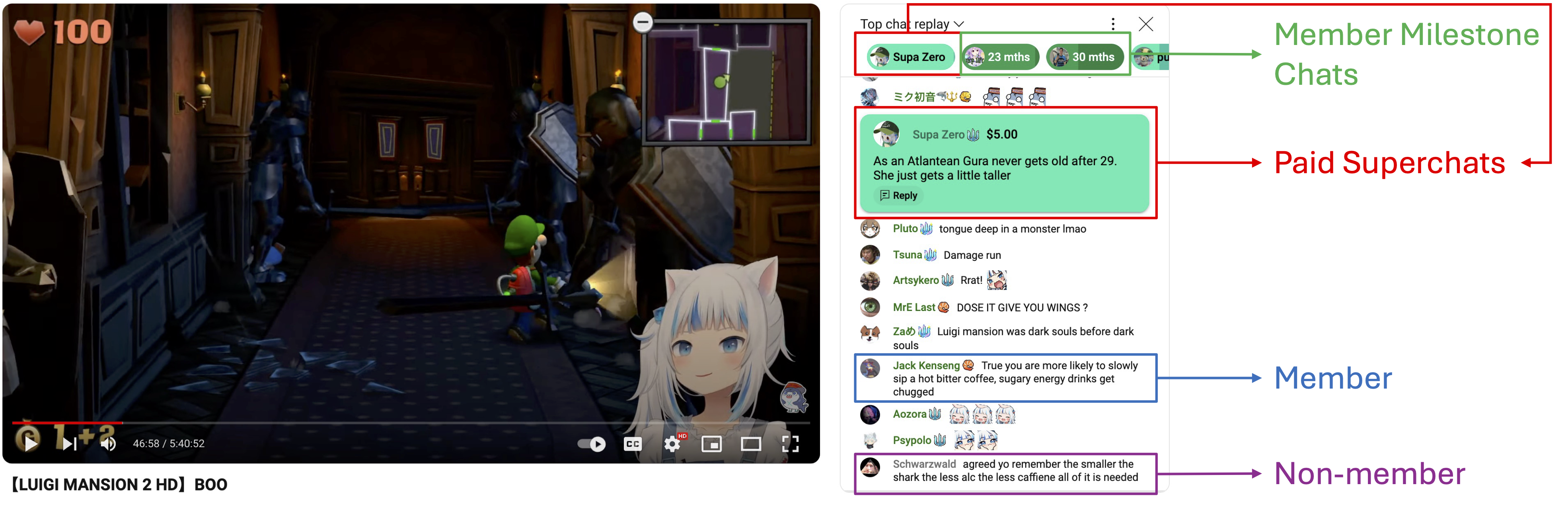}
    \caption{Screenshot of video game streaming by the Hololive VTuber \textit{Gawr Gura}, the most subscribed VTuber on YouTube \cite{gawr_four_million}.}
    \label{fig:livestreaming_interface}
\end{figure*}

\section{Introduction}

\textit{Virtual YouTuber}, or \textit{VTubers}, are a special category of livestreamers with virtual 2D or 3D anime-style avatars controlled by human voice actors. Under the masquerade of virtual bodies, \textit{Nakanohitos}, or the human voice actors can flexibly perform a constructed personality that appeals to their fanbase \cite{lu2021more,wan2023investigating,schmieder2024waiting}. Despite their virtual appearance, VTubers attract and engage a large fanbase through real-time interactions that provide emotional and entertaining affordance \cite{chinchilla2024vtuber}. Meanwhile, the audience can not only interact with the VTuber through the text-based livechat channel, but also support the VTuber with Superchats or monthly memberships, which grant them privileges like pinned messages for a longer duration, exclusive emojis, and username highlights (see Figure~\ref{fig:livestreaming_interface}). 
Unlike mainstream YouTubers who regularly upload pre-recorded videos and monetize through advertisement revenues \cite{kopf2020rewarding}, VTubers engage with the audience primarily through synchronized livestreaming and mostly monetize through Superchats and membership subscription.  
An online survey of 34 VTubers \cite{bilal2024vtuber} suggested that, among all monetization methods, Superchat donations (39.1\%) and membership subscriptions (26.1\%) are the two most popular ways. Financial report from Hololive Production \cite{cover2023} also suggested that Superchats and Membership were the most important revenue stream until 2022, after which they were only second to revenues from merchandising. 

VTubers initially debuted in Japan and became popular in East Asia \cite{tambunan2023factors}. In recent years, VTubers have also gained increasing popularity in the English-speaking cybersphere and worldwide, as evidenced by the English-speaking VTuber, \textit{Gawr Gura}, who reached over 4 million YouTube subscribers in about two years \cite{gawr_four_million}. As of May 2024, 64 out of the top 100 most Superchatted YouTubers are VTubers \cite{playboardMostSuper}, which collectively generate tens of millions in revenue per year. Despite their significant role in the content creation economy, the monetization landscape of VTubers are generally underexplored from the current literature.  


One distinction between VTubers and conventional game streamers is the companies behind them. VTubers, since their earliest days, are invented, manufactured and actively promoted by corporate agencies for commercial interests \cite{tan2025can}. VTubers also explicitly incorporate agencies in their self-branding \cite{marx2012jimusho}. Agencies, similar to `guilds' or Multi-channel Network (MCN) companies in the Western and Chinese livestreaming industry, are companies offering streamer support ranging from talent selection, content creation to monetization but also reaping a sizable portion from streamer incomes \cite{marx2012jimusho,zhang2019virtual,liang2024manufacturing}. Agencies have emerged as a critical player in the content creation economy, in addition to creators and platforms \cite{liang2024manufacturing}. Yet they have not received sufficient attention from researchers. Studying the VTuber community provides a unique opportunity to delve into these agencies, their economic activities, and their influence on VTubers and viewers. This will not only reveal how the VTuber community functions but also provide insights into how content creation has evolved.

In this study, we aim to understand the monetization of VTubers and the agencies behind them from a quantitative perspective by answering the following specific research questions (RQs):  

\begin{itemize}
    \item \textbf{RQ 1: Who profits the most from the VTuber livestream market?}
    \item \textbf{RQ 2: Who are contributing to the VTuber's monetization during livestreaming?}
    \item \textbf{RQ 3: What is the influence of agencies in monetization?}
    \item \textbf{RQ 4: What factors affect the survival of VTubers?}
\end{itemize}

We conducted a comprehensive quantitative analysis of over 1 million hours of livestream records from more than 1,900 VTubers on YouTube over six years. Our main contributions are as follows:  
\begin{itemize}
    \item First, our large-scale analysis offers a global quantitative characterization of the monetization landscape of the VTuber community, showing that the financial inequality and the widening wealth gap in the VTuber community, and that a subset of viewers contributes disproportionally to the incomes during livestreaming. 
    \item Secondly, from the perspective of agencies, we show that two agencies, Hololive and Nijisanji monopolize the VTuber market by owning the most profitable VTubers and that the dominance of these two agencies has been strengthened over the years. These two large agencies had a higher percentage of members and Superchat senders than VTubers from small agencies and independent VTubers, but a smaller percentage of loyal viewers. Agencies also cultivated a loyal fanbase loyal to the brand of the agency, rather than only to individual VTubers. 
    \item Furthermore, our survival analysis indicates that VTuber lifespans are highly associated with economic factors and, in particular, affiliation status. VTubers from large agencies survive longer than those from small agencies and independent VTubers. Yet VTubers from small agencies were active for a shorter lifespan than independent VTubers. 
    \item Based on our analyses, we discuss the increasingly critical role of agencies, or MCNs, in the creator economy and its implications. We also propose the design implications for the VTuber community to expand the monetization strategies and provide more equalitarian opportunities to small VTubers.
\end{itemize} 
For replication, we release our code for scraping and data analysis, as well as some preprocessed data, at \url{https://github.com/brdiep113/chi-livechats}.

\section{Backgrounds and Related Work}
Our study is motivated by prior research on VTubers and the economic status of digital content creators. In this section, we summarized the related works. 

\subsection{VTubers as a cultural phenomenon}
VTubers are a relatively new phenomenon that has only attracted attention from the HCI and CSCW community in the past few years. 
Livestreaming, especially game streaming, is mostly male-dominated \cite{doring2019male,freeman2020streaming}, yet VTubers are firmly dominated by female voice actors because of their origins in the Japanese idol industry \cite{wan2023investigating}. Some VTubers have been drawing intense criticisms of misogyny and gender biases are common in VTuber communities \citep{chen2024host,doring2019male,lu2021more,sutton2021gendered,brett2022we,byron2023new,wan2023investigating}.  
However, the VTuber streaming community is becoming more diverse in terms of language and gender \cite{lu2021more,sutton2021gendered,brett2022we}. 

Existing studies have investigated how VTuber perform various personalities visually and vocally \cite{ferreira2022vtuber,bredikhina2022becoming,wan2023investigating,chen2024host,chinchilla2024vtuber}. Other studies on VTubers also examine user engagement with virtual avatars \cite{lu2021more,stein2022parasocial,xie2023all,mou2023good},  viewer perception \cite{nissen2023you}, and motivations \cite{lu2021more,choudhry2022felt}. Many studies also explore the design and the technical aspects of virtual livestreaming \cite{zhou2024timetunnel,drosos2022design,kanamori2023practical,xu2021research,chen2024conan,tang2021alterecho,nguyen2023turning}.
Prior studies, primarily based on qualitative interviews, provide an in-depth description of individual VTuber and viewer experience \cite{lu2018you,lu2021more,choudhry2022felt,bartolome2023literature,wan2023investigating}, but scant attention has been paid to the financial ecosystem of VTubers, as well as the collective patterns of VTuber monetization activities. Unlike many individual-based streamers, many prominent VTubers are associated agencies \cite{lu2021more,tan2025can}. Agencies are key players in the market as they actively manufacture, promote, and monetize VTubers. Yet exactly how agency affiliation affects the performance of VTubers has not been systematically investigated. 

VTubers, like other online influencers, actively mobilize intimacies with followers to forge parasocial relationship from fans \cite{abidin2015communicative,chen2016forming,hair2021friends}. It has been found that the strength of the parasocial relationship on social media predicts content consumption and purchase intentions \cite{labrecque2014fostering,rubin2000impact}. Such parasocial interaction can be drawn upon to maximize monetization and could potentially lead to overspending \cite{stephens1996enhancing,masuda2022impacts}. In this study, we operationalize parasocial attachment as either being financially or behaviorally committed to VTubers. We compare how these committed viewers differ from the rest of the viewers in terms of membership subscriptions and Superchat donations.  

\subsection{The economy of social media content creation}
In recent years, online content creation has emerged as an increasingly vital component of the entertainment industry. Understanding these digital patronage platforms can help communities design for and support content creators and contributors for an enhanced experience. Prior studies have explored the creator economy on platforms like OnlyFans \cite{vallina2023cashing}, YouTube \cite{andres2023youtube,hua2022characterizing}, Patreon \cite{regner2021crowdfunding,el2022quantifying,hair2022multi} and the financial ecosystems on different platforms \cite{bonifacio2020digital,hara2018data}. 
Other studies investigate the online consumption behaviors of users, particularly user churn and user loyalty \cite{hamilton2017loyalty,el2022quantifying,lee2023ju}. 
Content creators are also actively employing multiple strategies to achieve monetization from users \cite{tafesse2023content,ma2023multi}. 

VTubers' monetization falls into the category of a fan-based model \cite{hair2022multi}, in contrast to the advertisement-based model that dominated Internet content creation in the past decades \cite{evans2009online}. Unlike other content creators who rely on revenues from advertisement agencies, VTubers generally receive monetary rewards directly from the voluntary donations of fans. Different from many YouTubers, most VTubers choose to livestream rather than upload pre-recorded videos. Superchats sent during livestreaming, in addition to any regular monthly subscriptions, constitute a primary source of income for VTubers \cite{bilal2024vtuber,cover2023}. While Subscription-based monetization \cite{vallina2023cashing,el2022quantifying,hair2022multi} and online gifting in game streaming \cite{wohn2020live,zhu2017understanding,yu2018impact,li2021virtual,tu2018earning} have received extensive research, online gifting (Superchats) and subscription (membership) in the context of VTubers have not received enough attention relative to its proportion in creator economy \cite{zhan2023exploring}.


\subsection{Multi-media networks in digital content creation}
With the rise of digital content creation, MCNs has already become another major player in addition to content creators and platforms. MCNs originated in YouTube, and were utilized to professionize content production on the platform \cite{cunningham2016youtube,lobato2016cultural,hou2019social,zhang2024contesting}. Yet YouTube only treated MCNs as transitional tools in its commercialization process and began limiting the operations of MCNs \cite{zhang2024contesting}. While MCNs are not quite prominent in Western streamers, they are the primary business model in the Chinese livestreaming industry. Most current research on MCNs focuses on China, which has the largest MCN market in the world \cite{zhang2024contesting}. MCNs have been mentioned in early interviews of Chinese streamers \cite{zhang2019virtual,wang2019love}, noting their role in providing support for streamers but also taking away profits and autonomy from them. Recent research suggests that MCNs has gradually evolved from intermediaries merely facilitating content creation \cite{lobato2016cultural,hutchinson2023digital} to gatekeepers manufacturing influencers in bulk and strategically controlling content creation, becoming increasingly dominant \cite{liu2023zhibo,zhang2023invented,liang2024manufacturing}. 

Within the VTuber contexts, MCNs are usually referred to as \textit{Jimusho}, or agencies. The origin of VTubers also started from the effort to manufacture online virtual idols with the existing workflows in the Japanese idol industry augmented with technology \cite{black2012virtual,tan2025can}. While functioning like other MCNs, VTuber agencies also inherit the Japanese idol agencies \cite{marx2012jimusho}. Prior studies on VTubers have briefly mentioned the role of corporate agencies in replacing VTuber voice actors at will \cite{lu2021more,wan2023investigating}, highlighting the controlling power of corporate agencies. Yet the role of corporate agencies has rarely been explored at a wider scale and outside the Chinese context. Given their increasing importance in institutionalizing digital content creation, 
our study aims to provide a nuanced understanding of the role of MCNs, or agencies in VTuber contexts (we use these two terms interchangeably), in the content creation economy.  

\subsection{Challenges in digital content creation}
The distribution of wealth is highly unequal among streamer communities with a long-tail distribution \cite{houssard2023monetization,pilati2024mirroring}, with top streamers getting much more wealth than those at the bottom. While such patterns regularly emerge from economic activities, researchers have exposed and addressed the inequalities and biases in digital platforms exacerbating such inequalities. Female and other marginalized streamers are more likely to be subject to the unfair community guidelines of the platforms \cite{kopf2020rewarding,kopf2024corporate}, moderator and algorithmic biases in content moderation \cite{jhaver2019did, ma2021advertiser,caplan2020tiered,kingsley2022give}, and constant harassment \cite{uttarapong2021harassment,tomlinson2024community,obreja2023toward,nguyen2024exploring}. Some studies also expose the precarious nature of the digital content creation professions and their work conditions \cite{de2011creative,morgan2018creativity,cheon2024creative}, highlighting how the nature of this profession differs from jobs in traditional sectors. It has been found that the actors animating virtual characters are facing vulnerabilities as employees, with their invisible labor not receiving sufficient recognition \cite{cheon2024creative}. 

Through the lens of quantitative analysis, we also plan to investigate the likelihood and the career of VTubers. As digital content creators, VTubers are likely to run into similar challenges such as job precarity, unstable incomes, and high dropout rates. Unlike most game streamers that are individual-based \cite{sjoblom2019ingredients}, the earliest and most successful VTubers were corporate-managed streamers \cite{bbcVirtualVloggers,scmpMeetJapans}. VTubers are also likely to face unequal power dynamics from corporate agencies or competitive pressure from the corporate content creation pipeline, which merits further research. 

\section{Method}

\subsection{Data collection}
For this study, we collected a large-scale dataset of livechats in VTuber streaming from YouTube. 

\subsubsection{Aquisition of VTuber samples} We first acquired a comprehensive list of VTubers from Virtual YouTuber Wiki\footnote{\url{https://virtualyoutuber.fandom.com/wiki/Virtual_YouTuber_Wiki}}, which covers around 2,800 VTubers streaming in more than 15 languages. VTuber Wiki had been used as a list for calculating the demographics of VTubers \cite{wan2023investigating}. This list, while not exhaustive, still constitutes a representative sample of the whole VTuber spectrum. 

\subsubsection{Meta information extraction} The fandom pages included detailed and structured meta-information of VTubers contributed by crowd-sourced editors. We used regular expressions to extract names, debut dates, agency affiliations (if any), channel links, presented genders, primary language, and (virtual) personal details. The extracted data were all manually inspected and corrected for erroneous information and typos. After inspecting, we selected only VTubers that had a YouTube channel link on their Fandom pages. 

\subsubsection{Livechat collection} Most VTubers archive all of their public streaming sessions with time-synchronized live chats. We used a customized scraper built on \texttt{yt-dlp}, an open-source YouTube downloader, to collect data from the channels of these VTubers. Given a link to a VTuber channel, the scraper retrieved all the public livestream records in the channel. Only text data for streaming sessions, including video descriptions and livechat records were collected. The video description included the title, the host, views, subscriber count, and duration. Each livechat session was a JSON-formatted object of all livechats sent in the livestreaming session, each livechat represented as a single entry with features listed in Table~\ref{tab:livechat_summary}. 

\subsubsection{Ethical practices} During data collection, we adhered to ethical practices for web scraping. To avoid overloading the website server, only necessary data was accessed and data requests were sent at a reasonable interval (> 5 seconds per request) to minimize high-frequency requests. The web scraping was completed in around four months with multiple breaks every day. 
The data collection procedure did not involve any direct interactions with the viewers or the VTubers. Hence this work does not fall into the categories of human subject research and no institutional review is needed.
All of the collected livestreaming sessions were public and non-sensitive. Users were fully anonymized through non-reversible hashing. We did not anonymize the VTubers as they are public figures.

\begin{table}[]
\small
\begin{center}
\begin{tabular}{|p{0.2\linewidth}|p{0.7\linewidth}|}
\hline
\textbf{Items} & \textbf{Description} \\ \hline
Session ID & The unique identifier of the livestreamed session\\
User ID & The public username, which we anonymized through hashing \\
Video Offset & The timing of the livechat relative to the video start time \\
Message & The text content of the message, including stickers \\
Superchat & Whether this livechat is a Superchat or not \\ Unix Timestamp & The Unix timestamp of the livechat \\
Author Badge & Whether the sender is a member at the time and, if yes, the duration in month \\
Purchase Amount & The precise value of Superchat in local currency, such as \$5, C\$5, NT\$100, \textyen 1000. \\
Message Duration & Duration of the message as proportional to the Superchat value \\\hline
\end{tabular}
\end{center}
\caption{A list of features available in each public livestream session record retrieved from YouTube.}
\label{tab:livechat_summary}
\end{table}

\subsection{Dataset summary}
Table~\ref{tab:summary} presents the overall summary of VTubers in our data. After data cleaning, we aggregated livestreaming sessions from 1,923 VTubers between early 2017 and July 2023, representing VTubers more than 15 languages. Altogether, these VTubers streamed more than 1 million hours in around 450k sessions on YouTube, yielding more than 17 billion views in total. These statistics suggest that our data represents a comprehensive view of the VTubers on YouTube. The VTuber landscape is firmly dominated by corporate agencies, as the two largest agencies, Hololive and Nijisanji, reap more than 60\% of total views. Most VTubers are affiliated with agencies, yet independent VTubers also contribute to a sizeable portion of streamed content.

\begin{table*}[]
\small
\begin{tabular}{lccccc}
\toprule
Affiliation & Num. VTubers & Total Videos & Total Hours & Total Views & Total Chats  \\\midrule
Affiliated & 1238 & 337k & 893k & 16324m & 9728k \\
\hspace{0.2in}- Hololive & 70 & 43.3k & 116k & 6330m & 3753k\\
\hspace{0.2in}- Nijisanji & 192 & 105k & 294k & 7091m & 3440k\\
\hspace{0.2in}- 774 Inc & 25 & 20.4k & 43.2k & 486m & 337k \\
\hspace{0.2in}- Other Affiliations & 951 & 169k & 441k & 2417m & 2199k \\
Independents   & 635 & 108k & 273k & 1007m &	723k  \\
\bottomrule         
\end{tabular}
\caption{Data for VTubers by Most Popular Affiliations.}
\label{tab:summary}
\end{table*}

\subsection{Extracting financial transcations}
We focus on two primary approaches to achieving monetization directly during livestreaming, \textbf{Superchat} and \textbf{Membership}, as illustrated in Figure~\ref{fig:livestreaming_interface}. 

\textbf{Superchats} are messages or stickers that will stay highlighted in bright colors and special fonts for a longer fixed interval on the livechat interface. 
The duration of the Superchats is proportional to the amount of payment. The price of Superchat is customizable between \$1 to \$500, though the actual price range varies across international regions and currencies.

\textbf{Membership} is another common way for fans to pledge their loyalty. A user can pay a monthly subscription fee to become a channel member, which, in return, receives member-only benefits including highlighted usernames, exclusive emojis, member-only livestream, and more close interactions with the VTuber. The membership also comes with different tiers, with the most common pricing ranging from \$1.99, \$4.99, \$14.99 to as high as \$34.99. 

In addition to Superchats and membership, other possible ways of monetization include streaming sponsored content, selling VTuber-themed products, serving as brand ambassadors, live concerts, etc \cite{tan2025can,cover2023}. However, these activities do not directly happen during livestreaming and the resulting transaction information is usually not disclosed. Therefore, we limited our analysis to the publicly available Superchat and membership information.

\subsubsection{Extracting membership}
We only counted the number of members as well as the maximal duration of membership per VTuber channel. Each member is associated with a member-only badge, which are usually marked with time spans such as one month, three months, and one year. For each member, we aggregated all their membership badges across all sessions for each VTuber and treated the longest period as indicated in their badge as the membership duration. The membership tiers are not differentiated in the public data, so we had to make a simplifying assumption that all membership here is homogenous.

\subsubsection{Extracting Superchats}
Each Superchat is associated with a numerical value in local currency as well as the currency symbol. A total of 61 unique currencies were identified through regular expressions. All payments were converted to US dollars with the daily foreign exchange rates available from the European Central Bank and \texttt{investing.com} (for 30 currencies not available in the European Central Bank). We used the exact timestamp of each Superchat to retrieve the daily currency exchange rate on the day when the Superchat was sent. 
We also manually verified the obtained numbers and rankings with public VTuber statistics\footnote{\url{https://playboard.co/en/youtube-ranking/most-superchatted-v-tuber-channels-in-worldwide-total}} and concluded that the numbers were indeed close, despite the fact that these public statistics also include private livestreaming sessions and deleted sessions.

\subsection{Data coding}
\subsubsection{Categorization of viewers}
\label{sec:loyalty}
Our viewer population only includes the viewers who left at least one livechat in livestreaming. We distinguished viewers who were committed behaviorally (loyal vs. non-loyal) and who were committed financially (Superchat senders and members). In previous literature, user loyalty is considered a strong proxy for latent engagement in online community platforms \cite{hamilton2017loyalty}. 
Adapted from prior work \cite{el2022quantifying, hamilton2017loyalty}, we adopt a time-dependent approach to defining \textbf{loyal viewers} by incorporating both users' preferences and commitment. Preference refers to users being active in a substantial portion of a specific VTuber's livestreaming sessions, while commitment refers to the sustainability of this preference over time \cite{hamilton2017loyalty}. Specifically, for a VTuber, a viewer is considered loyal in month $\mathbf{t}$ if they have sent at least one chat in at least half of this VTuber's streams in both month $\mathbf{t}$ and month $\mathbf{t+1}$.  For financially committed viewers, \textbf{Superchat senders} are those who have sent Superchats at least once. \textbf{Members} are those who pay to become a channel member at least once. Notably, a viewer can simultaneously exhibit behavioral loyalty and financial commitment to a VTuber (e.g., a viewer can be both a loyal viewer and a Superchat sender in month t). In our analysis, we examine viewer engagement and financial transactions separately. 

\subsubsection{Categorizing VTubers}
We also coded VTubers into three categories. VTubers themselves are not a homogenous community, agency affiliations can impact their actual activities \cite{tan2025can}. VTubers without any affiliations were coded as \textbf{Independent} VTubers. For affiliated VTubers, they were categorized into \textbf{Large Agency} VTuber if they were affiliated with Hololive and Nijisanji. Otherwise, they were coded as \textbf{Other Agency} VTuber. This categorization scheme allows us to compare the performance of affiliated and independent VTubers. Hololive and Nijisanji VTubers were coded as separate categories as these two companies were outstanding from other agencies in terms of various metrics. 

\subsubsection{Constructing VTuber networks through loyal viewers}
Viewers often follow more than one VTubers. To visualize this, we constructed the loyal viewer networks as a weighted undirected graph between VTubers, with each VTuber as a node. Let $A$ and $B$ be two lists of loyal viewers from two VTubers. The edge weight between these two VTubers is the same as the dice coefficient of their loyal viewers, defined as:
\[
D(A, B) = \frac{2 |A \cap B|}{|A| + |B|}
\]
where $|A \cap B|$ is the intersection of loyal viewers and $|A|,|B|$ are the total number of loyal viewers. Dice coefficient quantifies the fraction of shared loyal viewers between VTubers. We computed the dice coefficient for each unique pair of VTubers and used them as the edge weights between this pair of VTubers. This network allows us to visualize the relationship between VTubers.

\subsection{Analyzing income disparity}
In order to quantify the monetary disparity between viewers, we calculated Gini coefficients, or Gini index. Gini index has been a common measure of economic inequality and has been applied to compare the income polarity in online platforms \cite{el2022quantifying,houssard2023monetization}. Given a sequence of $n$ values $\{x_1, x_2, \dots, x_n\}$ that is sorted in ascending order, the Gini coefficient can be calculated using the following definition \cite{dixon1987bootstrapping}.
\begin{equation}
G = \frac{\sum_{i=1}^{n}(2i-n-1)x_i}{n\sum_{i=1}^{n}x_i}
\end{equation}
where $i$ is the rank of $x_i$. G is the Gini index with a theoretical range between 0 and 1, where 0 implies absolute equality and 1 means extreme inequality. We used Gini index to quantify the monetary disparity in the following three scenarios. 
\begin{itemize}
    \item \textbf{Between-VTuber Gini index}. This Gini index measures the disparities of total income between each VTuber. 
    \item \textbf{Within-Vtuber Gini index}. This Gini index measures the income stability of a VTuebr by quantifying the Superchat income disparties across all livestreaming sessions for the VTuber. Membership subscription fees are not included in this calculation, as we do not have access to such data.
    \item \textbf{Superchat Gini index}. This Gini index quantifies the disparities in monetary contributions between Superchat senders for each VTuber. It is calculated based on the total contributions by Superchat sender for each VTuber. 
\end{itemize}

\subsection{Survival analysis}
We adopted a survival analysis to analyze the probability of VTuber failure given their monthly engagement features over time. Survival analysis is a collection of statistical methods concerned with predicting the expected temporal span until an event occurs \cite{kleinbaum1996survival,cox2018analysis}. In our case, the event of interest is when the VTuber will cease streaming.

\subsubsection{Feature extraction}
Our data is an incomplete snapshot of VTuber activities, as many of them will continue to remain active after the data collection cutoff. As for defining failure in our analysis, a VTuber is considered to have failed at their \textit{n}-th month if the time between the VTuber's most recent stream to their first stream is less than $n$ months. We also considered any VTuber's who's most recent stream to fall within the 3 months of the cutoff date to still be right-censored, implying that their survival time is potentially longer than the observed label \cite{kleinbaum1996survival}.

In our analysis, we considered time-dependent viewer engagement features at each month as well as demographic information such as affiliation and gender of VTuber. We also considered engagement features such as number of members, number of unique viewers, Superchat income, and duration streaming from every month from the date of their first stream to their failure or censoring. The set of all features are summarized in Table \ref{tab:feat_summary}. 
We ensure that our features can be used to predict only future failure by predicting the survival status of each VTuber at month $\mathbf{t}$ with features from month $\mathbf{t - 1}$.



\begin{table}[]
\small
\begin{center}
\begin{tabular}{|p{0.2\linewidth}|p{0.7\linewidth}|}
\hline
\textbf{Feature} & \textbf{Description} \\ \hline
Affiliation & The listed affiliation of the VTuber as listed on the VTuber Fandom Wiki. This is coded as a categorical variable with 3 levels, large affiliation for those affiliated with Hololive or Nijisanji, other affiliations for those with any other affiliations, and independent for those with none. \\ \hline
Gender & The listed gender of the VTuber as listed on the VTuber Fandom Wiki (including male, female, and other/non-binary) \\ \hline
Total Stream Duration & Total duration of each of a VTubers' recorded streams \\ \hline
Total Suprchat Income & Total amount of Superchats received (converted into \$USD equivalent) over each of a VTubers' recorded streams \\ \hline
Total Viewers & Total number of unique viewers who left comments on the VTubers streams \\\hline
Total Number of Members & Total number of unique users subscribed to the VTuber \\ \hline
\end{tabular}
\end{center}
\caption{Summary of all extracted features for each month.}
\label{tab:feat_summary}
\end{table}
\subsubsection{Cox model} The model we fit is a Cox proportional hazards model \cite{cox1972regression}. It is a parametric model that estimates a baseline hazard (risk of failure) function over time as well as the effect of additional covariates on the baseline hazard. Positive coefficients represent increases in the risk of failure, whereas negative coefficients represent decreases in risk of failure. Under exponentiation, the coefficients of the model can also be interpreted as representing a hazard ratio representing the multiplicative change in hazard, or the risk of failure. Notably, several of our covariates fail a test of proportionality \cite{grambsch1994proportional} at the $\alpha = 0.05$ significance level. This suggests the typical assumption that the affect of covariates on the hazard does not vary in time. In order to address this, we include interaction terms in our model with the covariates that violate fail the proportional hazards assumption to reflect the time-varying nature of this problem.  
All hazard ratios in our analysis are reported with female and large affiliation (Hololive/Nijisanji) VTubers as the reference.

We also estimate the predictive accuracy of our model by doing a 5-fold group cross-validation such that no samples of a VTuber exist in both the training and held-out fold. We calculate concordance index (Harrell's \textit{C}) on the held-out fold as our evaluation metric \cite{harrell1982evaluating}. Concordance index measures the proportion of all pairwise samples in the data where the model assigns higher risk for the subject with the shorter survival time. 


\begin{figure*}
    \centering
    \includegraphics[width=0.4\linewidth]{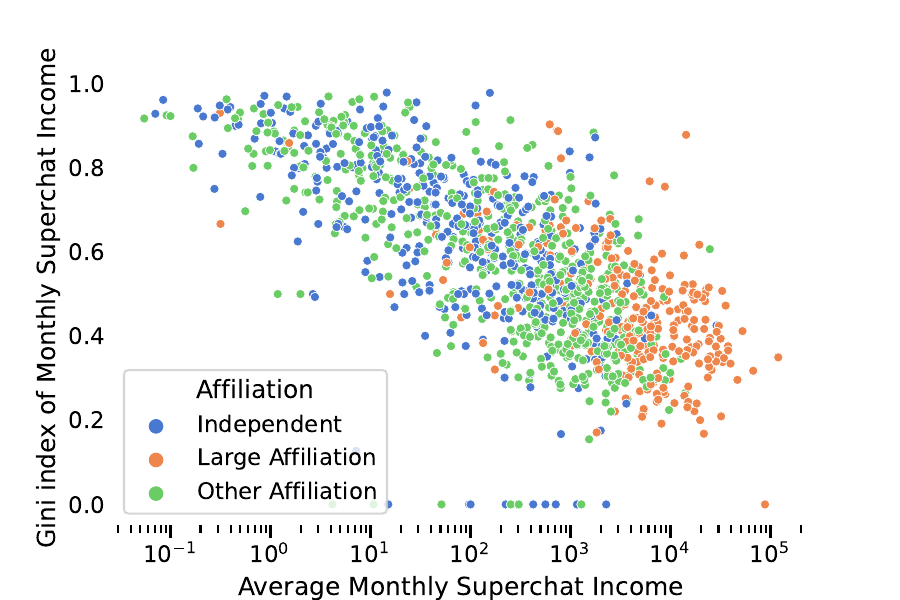}
    \includegraphics[width=0.3\linewidth]{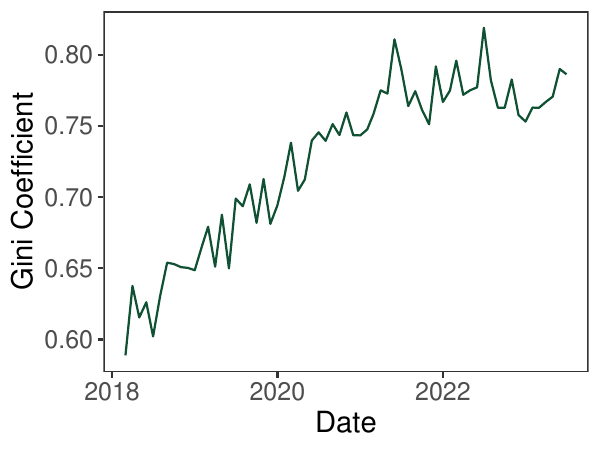}
    \caption{Left: VTubers' monthly Superchat Income and the Within-Vtuber Gini index of monthly income for each VTuber. A higher Gini index in this context suggests high variation of monthly income. Right: Between-VTuber Gini index across VTubers Over Time. The overall trend shows an increasing Gini coefficient from 2018 into 2023 when the data was collected. }
\label{fig:gini}
\end{figure*}

\begin{figure*}
\centering
\begin{subfigure}{.5\textwidth}
  \centering
  \includegraphics[width=0.9\linewidth]{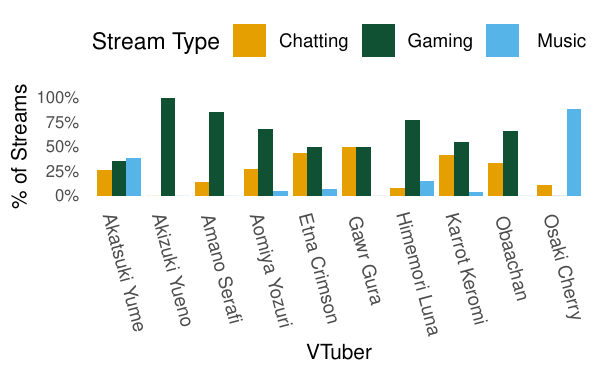}
  \caption{Distribution of Session Types.}
  \label{fig:top_streamtypes}
\end{subfigure}%
\begin{subfigure}{.5\textwidth}
  \centering
  \includegraphics[width=0.9\linewidth]{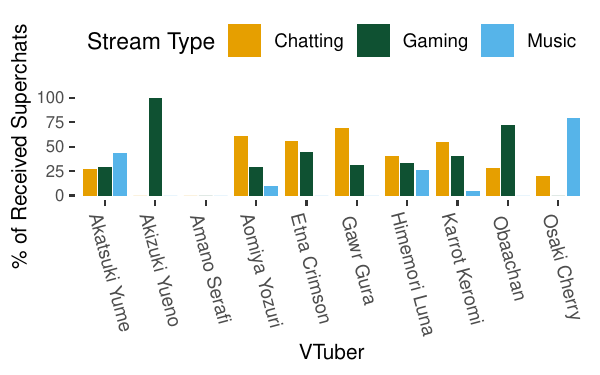}
  \caption{Stream Types by Superchats.}
  \label{fig:top_streamtype_incomes}
\end{subfigure}
\caption{Distribution of Stream Type. Many VTubers engage in multiple activities across their streams but gaming and chatting sessions are usually the most profitable.}
\label{fig:session_type}
\end{figure*}

\begin{figure*}[t!]
    \subfloat[Usada Pekora]{%
        \includegraphics[width=.4\linewidth]{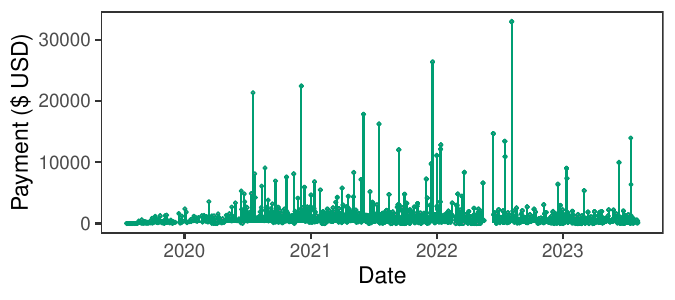}%
        \label{subfig:usada_pekora_income}%
    }
    \subfloat[Kamito]{%
        \includegraphics[width=.4\linewidth]{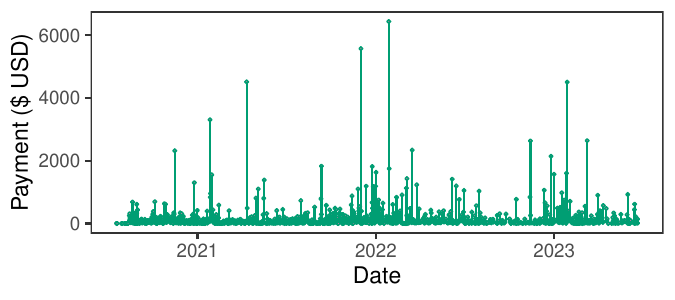}%
        \label{subfig:kamito_income}%
    }\\
    \subfloat[Gawr Gura]{%
        \includegraphics[width=.4\linewidth]{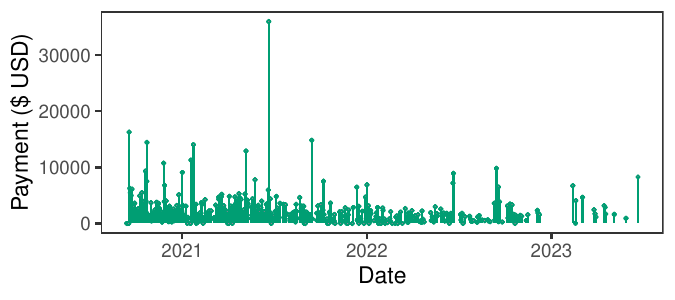}%
        \label{subfig:gawr_gura_income}%
    }
    \subfloat[Kson]{%
        \includegraphics[width=.4\linewidth]{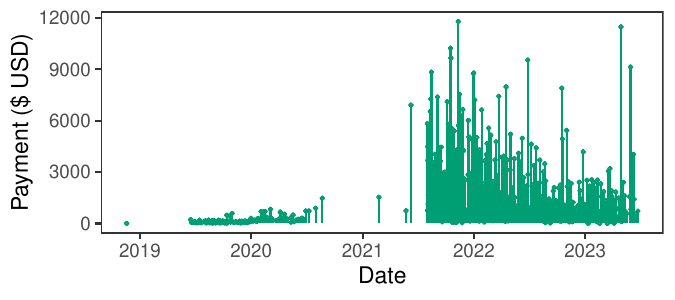}%
        \label{subfig:kson_income}%
    }
    \caption{Superchat incomes over time for selected VTubers. Income is unevenly distributed throughout a VTubers streams, with significant spikes in a few streams.}
    \label{fig:vtubers}
\end{figure*}

\section{RQ1: Who profits the most from the VTuber livestreaming market?}
In this section, we provide empirical answers to \textbf{RQ1}. Generally speaking, only a few top VTubers and agencies are profiting a lot from the livestream market. 

\subsection{The incomes of VTubers are highly unequal}
In Figure~\ref{fig:gini}, the between-VTuber Gini index  of VTuber earnings continued to grow from around 0.60 in 2018 to 0.75 in 2023, indicating a rapidly widening wealth gap between VTubers over time. VTuber incomes are highly concentrated, with only a small number of top VTubers reaping most fans and monetary earnings. 
Among all VTubers, only 75\% of them had received at least one Superchat, or one fourth of VTubers did not receive any Superchat within the investigated period. On average, the monthly income for all VTubers is \$2,667. Yet the income distribution is highly skewed, such that the median monthly income is only \$127. 

Figure~\ref{fig:gini} suggests that Gini indexes are low for VTubers with the highest and the lowest incomes. For VTubers with the lowest incomes, their monthly incomes are always close to 0. For highest-income VTubers, they have a large enough fanbase to sustain a high and stable income every month, and these VTubers are also more likely from large agencies. Generally speaking, higher income is weakly associated with higher income stability (Spearman correlation: $\rho = 0.30$, $p=2.82e^{-37}$), a trend also observed in Twitch streamers \cite{houssard2023monetization}.

\subsection{Most Superchat incomes are from regular gaming and chatting sessions (and a few special events)}

To understand what livestream content contributes to Superchats, we sample 10 VTubers from our dataset (ensuring representation from large corporate affiliations i.e. \textit{Hololive} and \textit{Nijisanji}, as well as smaller affiliated and independent VTubers) and manually tagged their streamed sessions from the first 3 months of 2023 to ensure consistency in time. We found that most livestream sessions rotate between gaming, chatting, and singing, though individual VTubers prioritize certain content types. In Figure \ref{fig:top_streamtypes}, a majority of streams are video gaming streams, more frequently than chatting and musical streams. Video gaming is popular among VTubers as it can offer a stable viewer count and is not as effortful. In comparison, singing requires special training and effort, so fewer VTubers are doing music streaming. However, in terms of earnings, gaming streaming generally earns fewer Superchats than chatting and music streaming. In Figure \ref{fig:top_streamtype_incomes}, chatting sessions tend to generate a higher percentage of Superchat values than gaming streams, despite being less frequent. In the case of the 324 sessions considered in this analysis, over 12\% (37 streams) were collaborated between two or more other VTubers from the same agencies. 
Many agencies often organize co-streaming events with several of their own VTubers \cite{lee2023effects}, which brings different viewer groups together, leading to a large number of Superchats. 

Figure~\ref{fig:vtubers} provides some qualitative examples of a few top VTubers' income per session over their lifespan. The Superchat income per session is relatively stable. Yet there are often some outlier sessions (spikes in figures) that yielded much higher incomes than average. These outlier sessions are usually special events such as birthday party concerts, collaborative live concerts, and new avatar reveal parties. These special events usually offer more novel content than average streams, thereby attracting more viewers and Superchats than regular sessions. A similar case has also been observed in Twitch that key events are effective ways to attract a larger-than-usual audience \cite{deng2015behind}.

\subsection{Superchats are mostly from a few key moments in livestreaming}
\begin{figure}
    \centering
    \includegraphics[width=\linewidth]{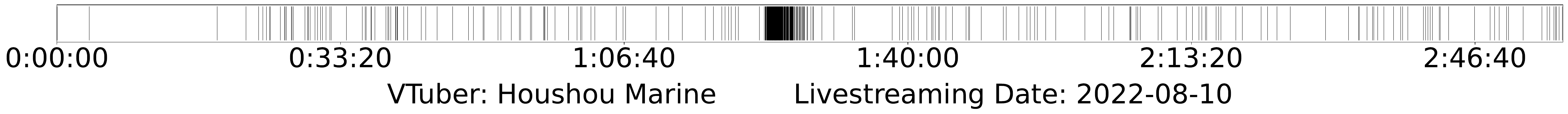}
    \includegraphics[width=\linewidth]{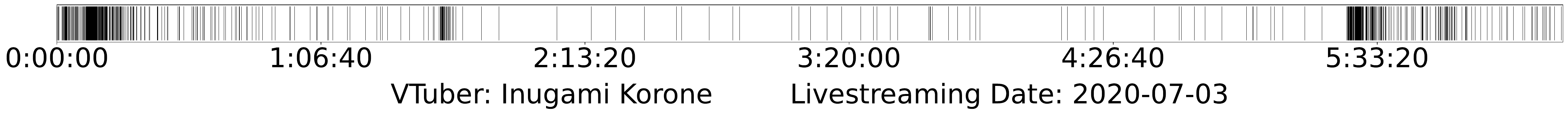}
    \includegraphics[width=\linewidth]{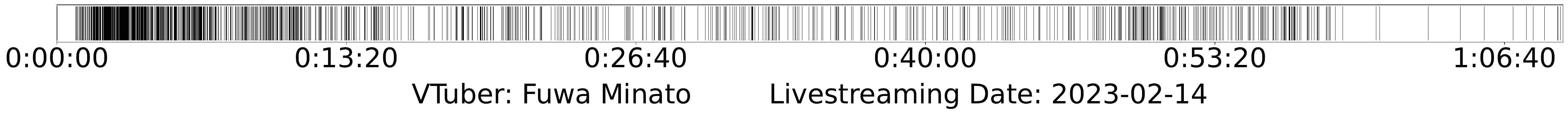}
    \includegraphics[width=\linewidth]{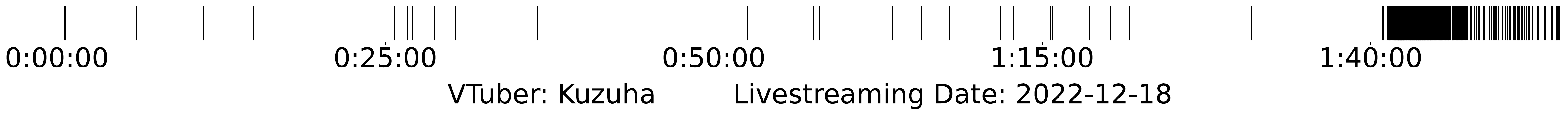}
    \caption{Temporal occurrences of Superchats across the whole sessions, with each dark bar indicating a Superchat. Superchats tend to occur in clusters rather than evenly distributed across sessions.}
    \label{fig:superchat_distribution}
\end{figure}

We randomly sampled 100 livestream sessions for qualitative analysis. Figure~\ref{fig:superchat_distribution} illustrates the temporal distribution of Superchats for selected livestream sessions. Consistent prior observation on gifting events \cite{zhu2017understanding, tu2018earning}, Superchats tend to be clustered in short time spans rather than evenly spread out across the whole session. Compared to the long-term commitment of membership, Superchats are more likely to be impulsive purchases, induced by the special moments in livestream and the behavior of peer viewers. 
Based on our analysis, there are three common cases where Superchats are sent in synchrony. First, Superchats are sent in bulks at the beginning and the end of the livestream, as a way of greeting. Secondly, most Superchats were sent out to celebrate the memorable, funny, or touching moments. The large immersive virtual concerts can attract tens of thousands of viewers, spurring more spikes of Superchats \cite{lee2023ju,lee2023effects}. 
Thirdly, in the case of conflicts or discussions, Superchats are sent out by viewers who truly want their voice heard by pinning their paid comment at the top of the chat window for a longer exposure.

\section{RQ2: Who are contributing to the VTuber's monetization during livestreaming?}

\begin{figure*}
\centering
\begin{subfigure}{.3\textwidth}
    \centering
    \includegraphics[width=1\linewidth]{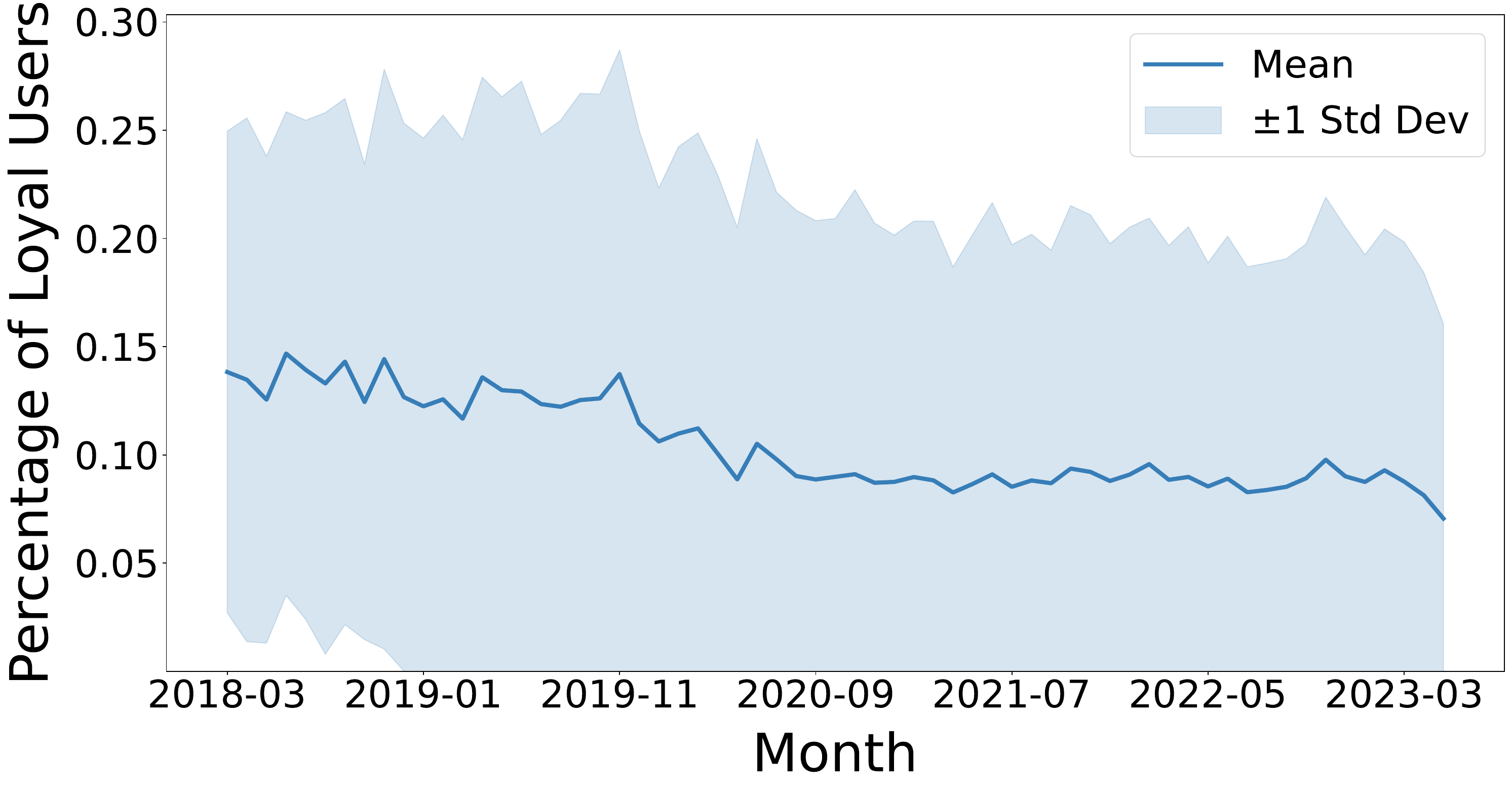}
    \caption{Average Percentage of Loyal Viewers across VTubers}
    \label{average_percentage_of_loyal_users_by_month}
\end{subfigure}
\begin{subfigure}{.3\textwidth}
    \centering
    \includegraphics[width=1\linewidth]{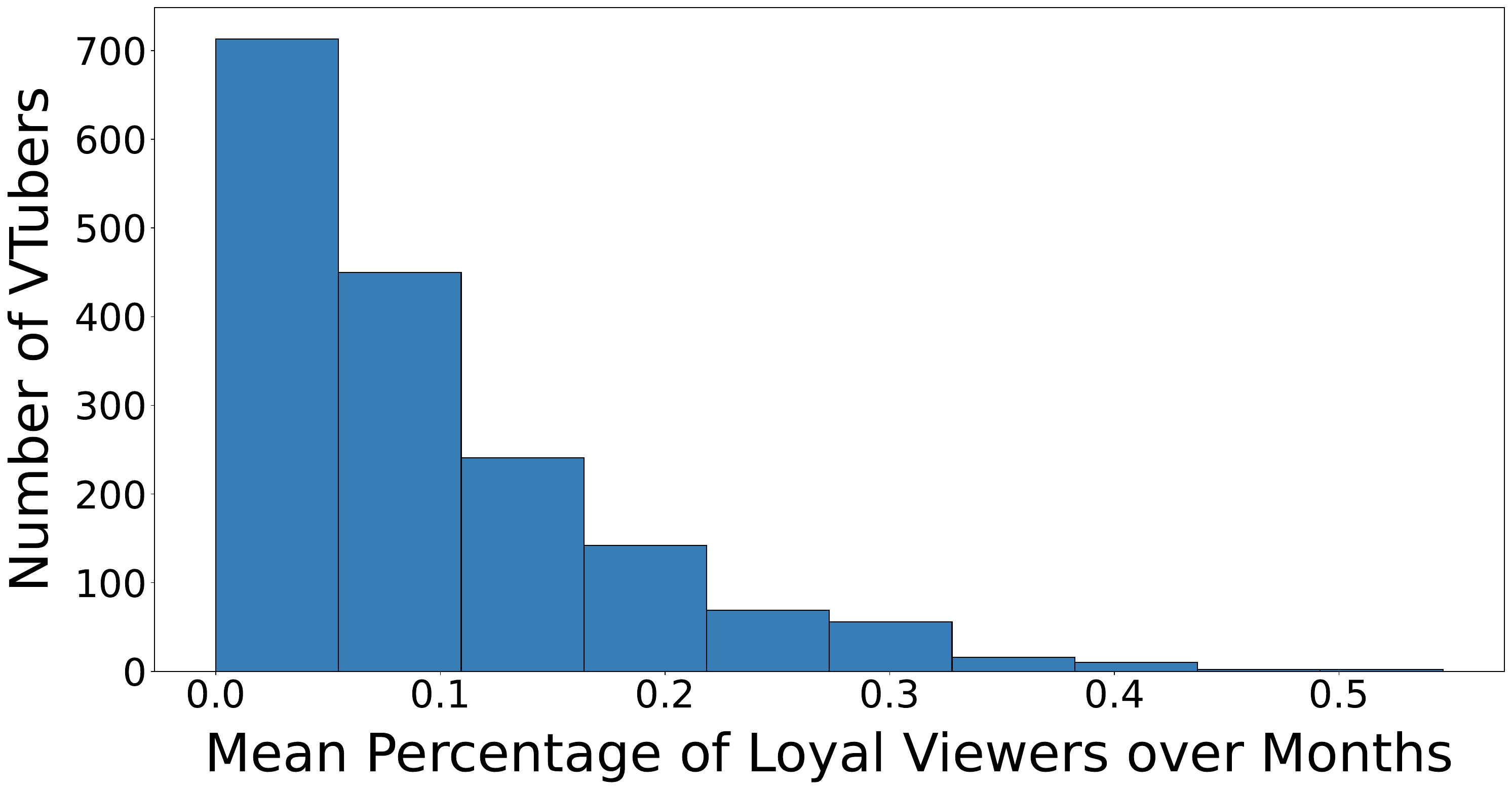}
    \caption{Average Percentage of Loyal Viewers across Months}
    \label{average_percentage_of_loyal_users_by_vtuber}
\end{subfigure}
\begin{subfigure}{.3\textwidth}
    \centering
    \includegraphics[width=1\linewidth]{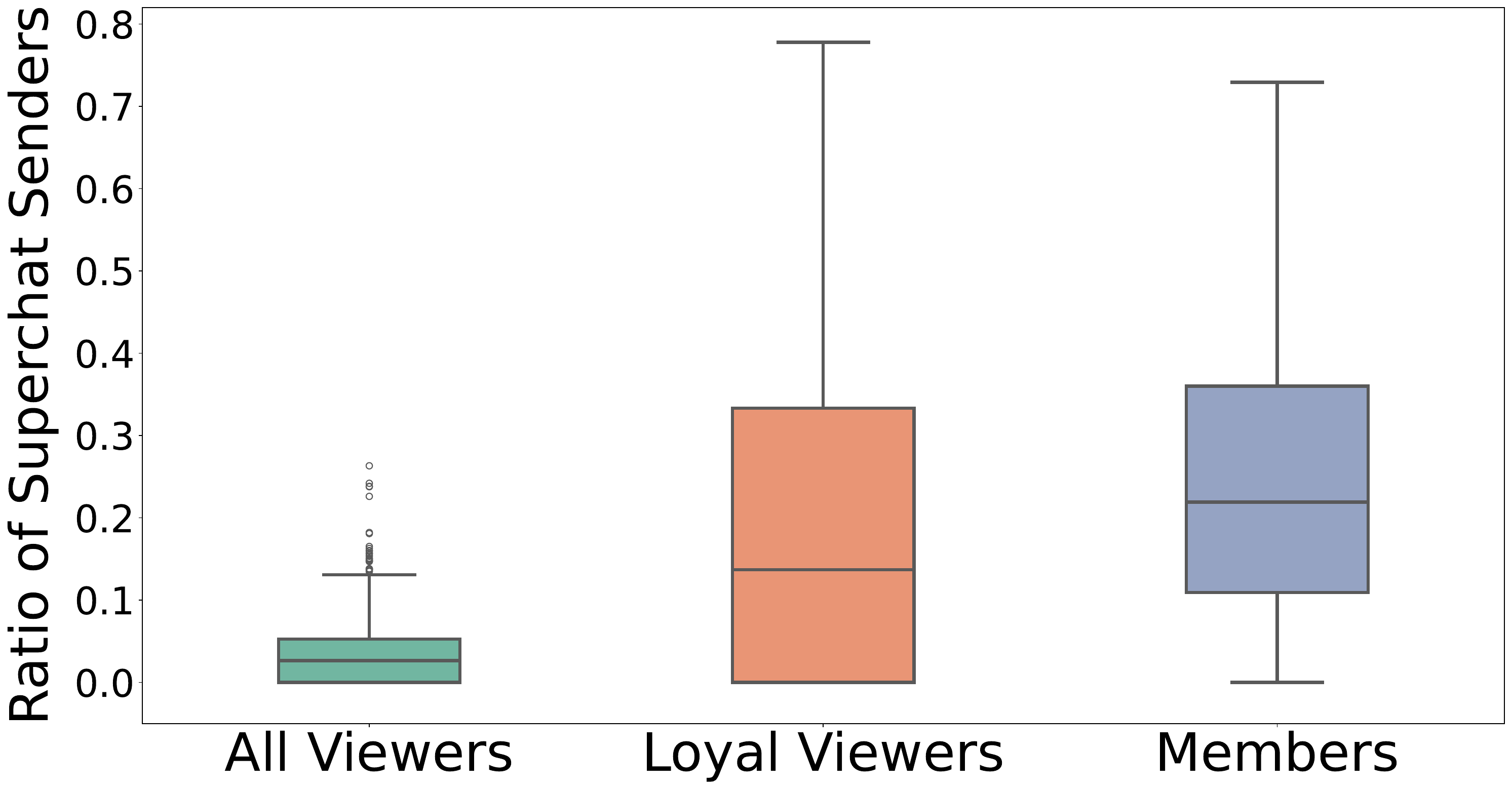}
    \caption{Distributions of Superchat Sender Ratios per Vtuber}
    \label{fig:superchat_senders_ratio}
\end{subfigure}
    \caption{Loyal Viewer Statistics. Kruskal-Wallis H-test revealed significant differences among the ratios of Superchat senders in all viewers, loyal viewers, and members ($H=1116.72, p=3.22\times10^{-243}$). Post-hoc analysis revealed significant differences in all pairwise group comparisons.}
    \label{fig:stats_of_loyal_users}
\end{figure*}

\subsection{VTubers' incomes are primarily from a small group of loyal fans}
We have aggregated 6.75 million unique viewers, among which 7.96\% have sent at least one Superchat and 12.65\% have purchased at least one VTuber membership. However, more than 80\% viewers in livechats did not spend any money. As we are only counting viewers who left at least one livechat, the actual ratio can be much lower if we count all viewers. Figure~\ref{fig:stats_of_loyal_users} indicates that the ratio of loyal viewers was generally low in the VTuber viewers. However, viewers who were behaviorally committed (loyal viewers) or financially committed (members) were more likely to send out Superchats than the general viewer population. 

\begin{figure}
    \centering
        \includegraphics[width=0.9\linewidth]{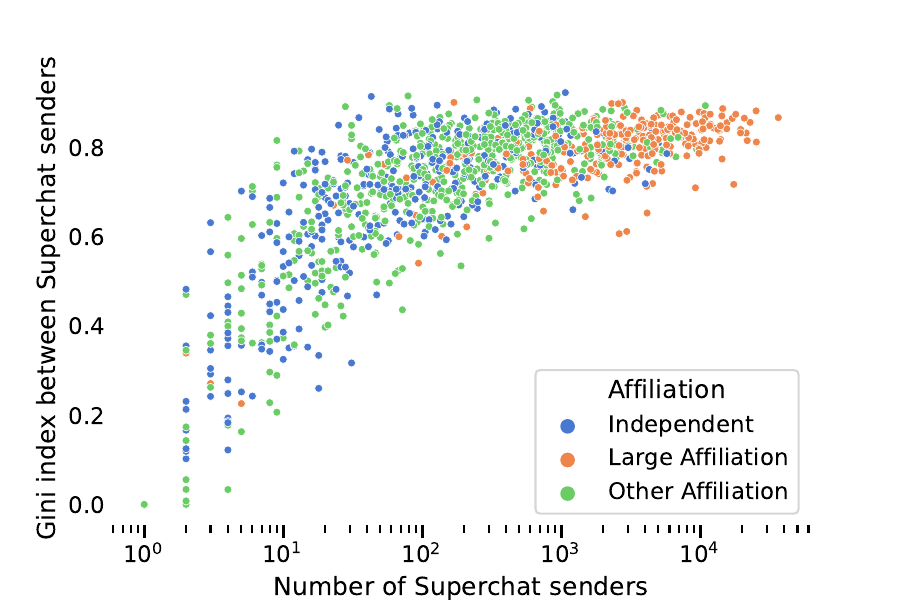}
    \caption{Superchat Gini index of Viewer Superchat Income Contributions. The Spearman correlation between number of Superchat senders and Gini Index is $\rho = 0.71$, with a $p$-value of $3.31 \times 10^{-191}$.}
    \label{fig:user_contribution_gini_index}
\end{figure}

Even among the Superchat senders, there was tremendous inequality in individual contributions. 
For most VTubers, their Superchat Gini indexes are around and above 0.6, indicating large inequality in viewer spending (see Figure~\ref{fig:user_contribution_gini_index}). This implies that the core contributing viewers are extremely small in number, yet contribute a disproportionally large share of income. Superchat Gini indexes also grow with the increase in the paying viewer base, but eventually saturates around 0.8-0.9. Generally, a larger paying user base implies more diversity in users across regions and economic conditions, which enlarges the monetary gaps. 


\subsection{Membership pledges tend to be short and exclusive}
Most VTubers did not set up membership subscriptions or did not have members, but for those who did, the average membership rate was around 5\%, slightly higher than the ratio of Superchat senders. Compared to the one-time Superchats that cost between \$1 to \$500 only for brief moments, the membership is a more cost-effective bargain to showcase loyalty. Not only because the price is cheaper, but also because members can get their usernames highlighted in color with an exclusive sticker attached, and receive additional highlights for Superchats for a whole month. 

\begin{figure*}
\centering
\begin{subfigure}{.35\textwidth}
    \centering
    \includegraphics[width=1\linewidth]{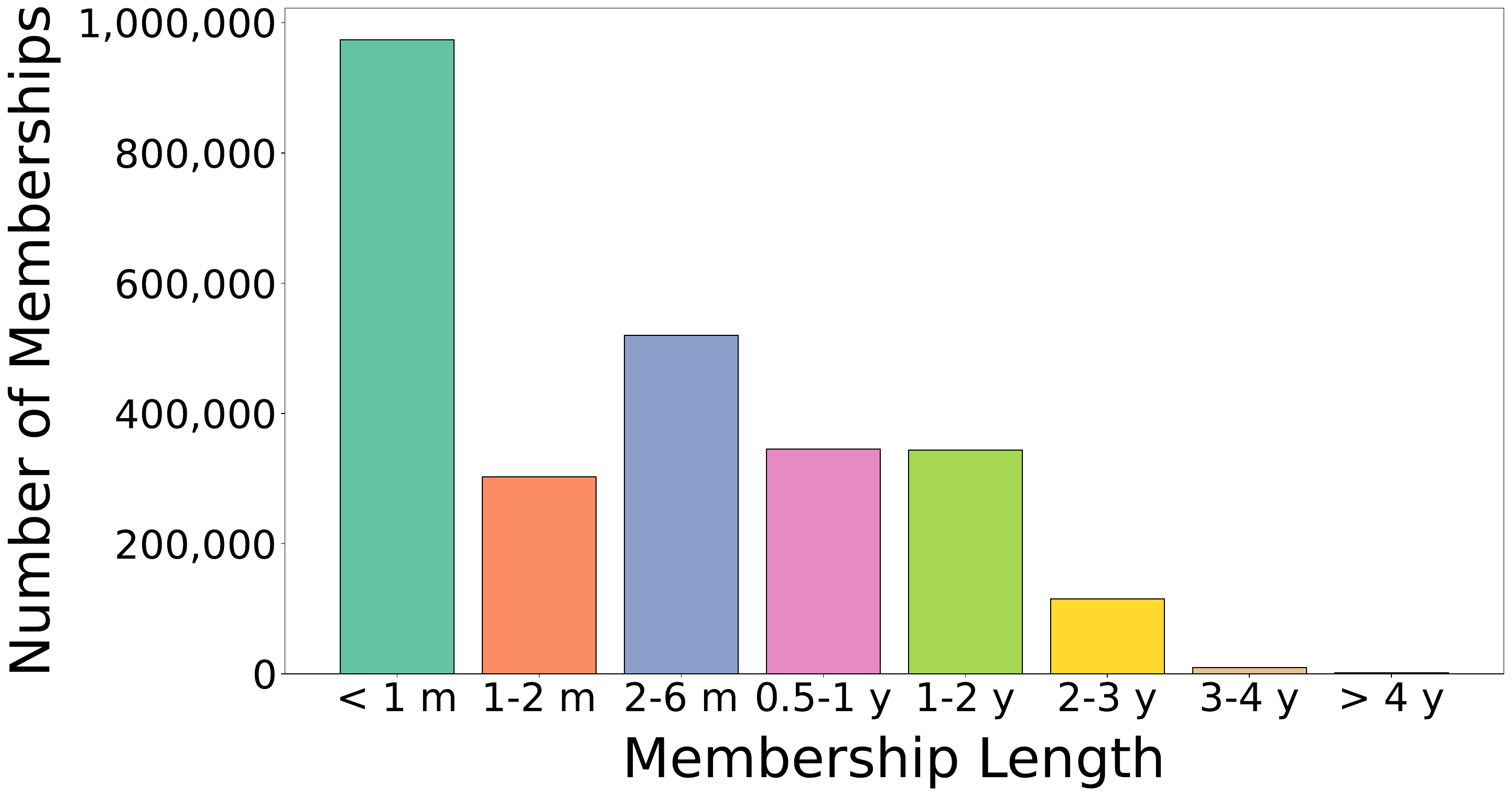}
\end{subfigure}
\begin{subfigure}{.35\textwidth}
    \centering
    \includegraphics[width=1\linewidth]{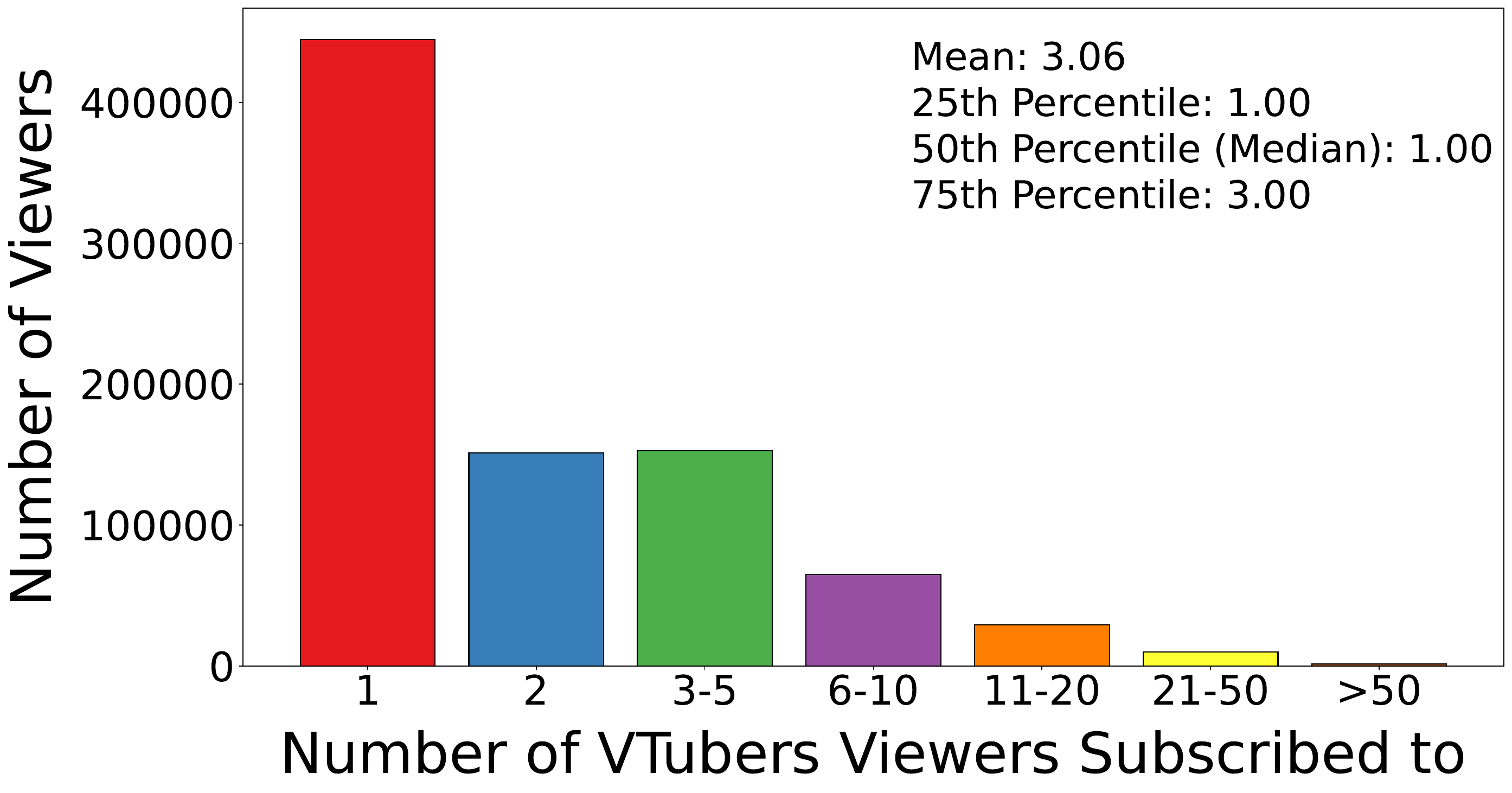}
\end{subfigure}
    \caption{Distribution of Membership Subscription Length; Number of Vtubers Viewers Subscribed to as a Member. The majority of users have a relatively low number of VTubers that they are subscribed to and tend to have short-term memberships.}
    \label{fig:membership_length_and_exclusivity}
\end{figure*}

Similar to many other content creators, VTubers are performing parasocial relational labor to attract viewers \cite{hair2021friends}.  Viewers' commitment to VTubers is usually exclusive. Our data analysis confirms that the majority of members (over 50\%) subscribed to only one VTuber throughout their active period (Figure~\ref{fig:membership_length_and_exclusivity}), as it can be difficult financially, and maybe in terms of time, to commit to multiple VTubers for most viewers.  Most membership also is also short-lived. In fact, over 40\% of members ended their membership in one month (Figure~\ref{fig:membership_length_and_exclusivity}). Yet, there is still a non-negligible portion of viewers who maintain their membership for more than 1 year, some even over 4 years. 
Such parasocial relationships can be hard to maintain over a long period.

\subsection{Superchats were mostly sent out in small amounts but loyal viewers sent them more frequently}

\begin{figure*}
    \centering
    \begin{subfigure}{0.33\textwidth}
        \centering
        \includegraphics[width=1\linewidth]{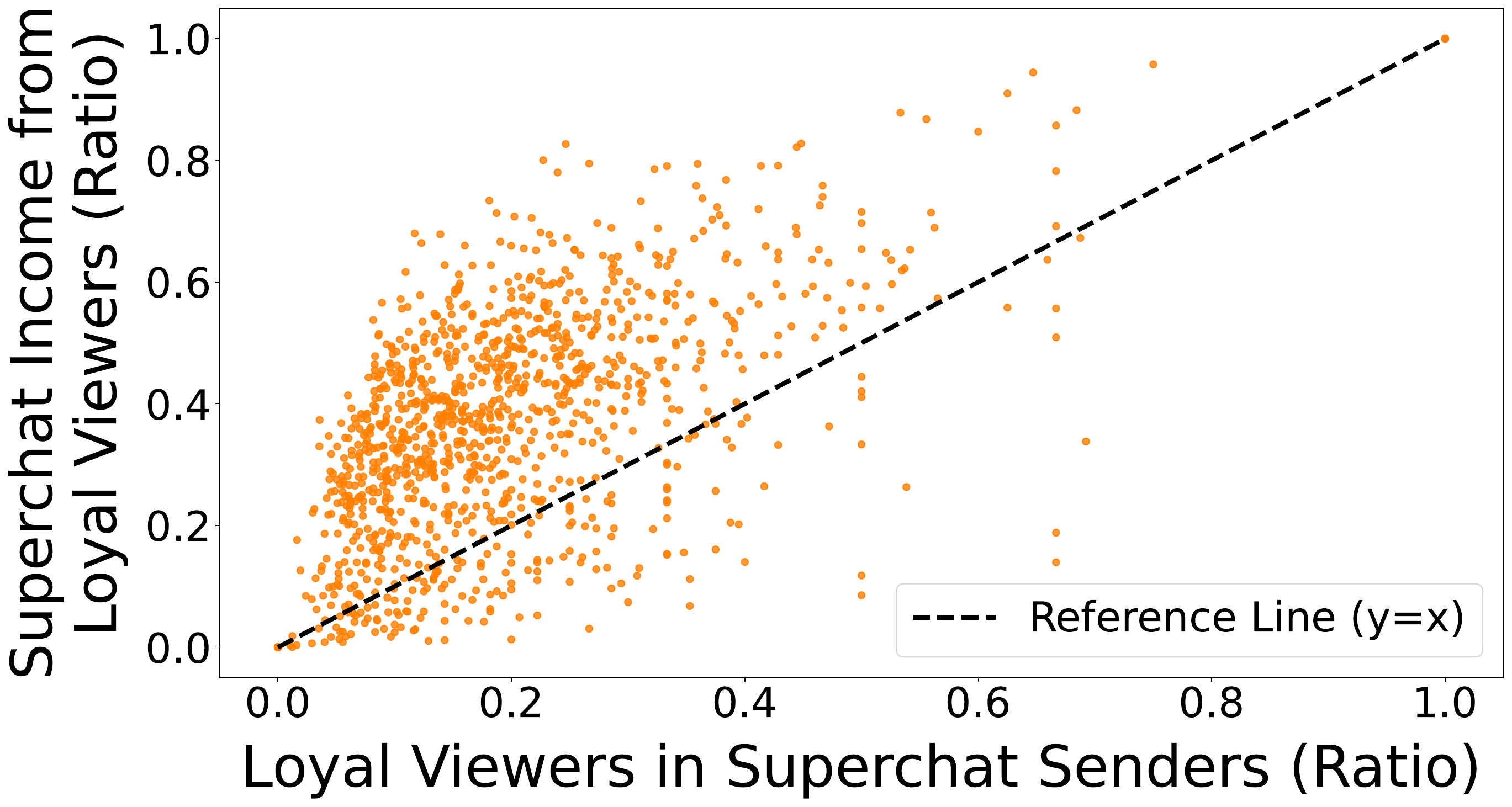}
    \end{subfigure}
    \begin{subfigure}{0.33\textwidth}
        \centering
        \includegraphics[width=1\linewidth]{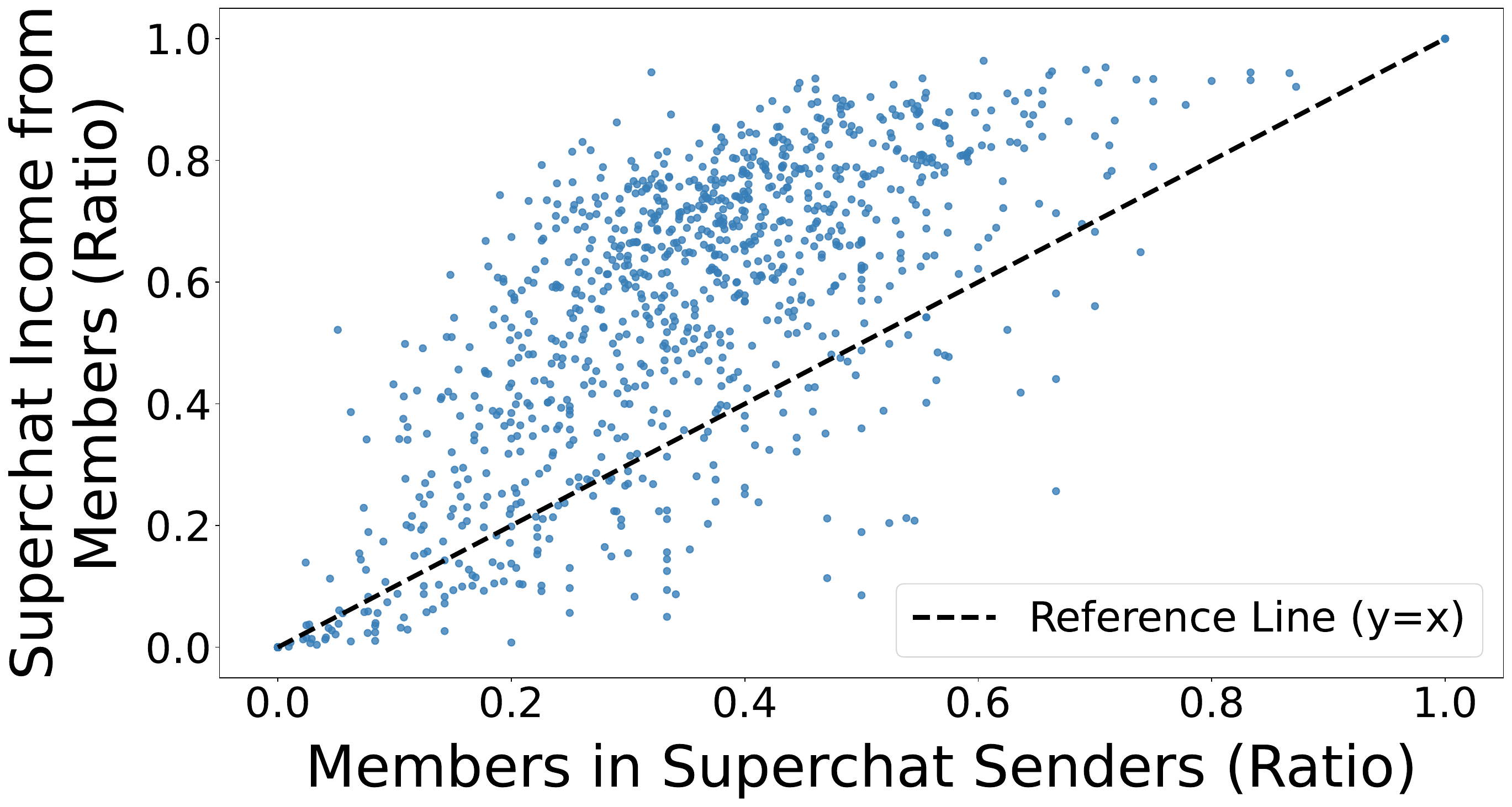}
    \end{subfigure}
    \begin{subfigure}{0.29\textwidth}
        \centering
        \includegraphics[width=1\linewidth]{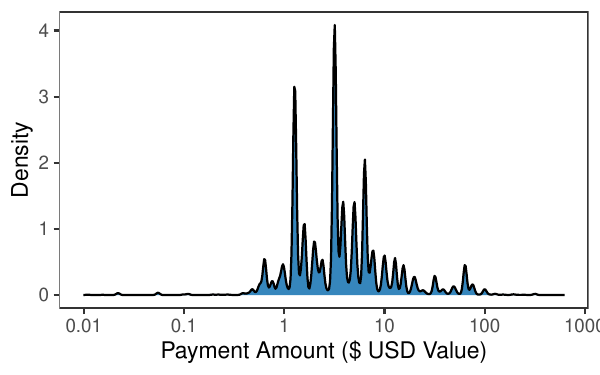}
    \end{subfigure}
    \caption{Superchat Income Contributions of Loyal Viewers and Members. Loyal viewers and members contribute disproportionately more monetarily.}
    \label{fig:scatter_plots_superchat_concentration}
\end{figure*}

The distribution of individual Superchat payments in Figure~\ref{fig:scatter_plots_superchat_concentration} suggests that most Superchats cost around \$1-\$10. The resulting probability distribution has multiple spikes because the payments are usually limited to whole number amounts. YouTube caps the maximum amount of Superchats to \$2,000 per week, so it is rare to find Superchats beyond \$500 in the data. 

We compared the Superchat contributions from loyal viewers and members among all Superchat senders in Figure~\ref{fig:scatter_plots_superchat_concentration}. If the ratio of loyal viewers or members is proportional to their percentage of contributed Superchat values, the data points in Figure~\ref{fig:scatter_plots_superchat_concentration} should fall exactly on the reference line. However, most data points are above the reference line, implying that they contributed unproportionally to Superchat incomes of the VTubers. 
Figure~\ref{fig:superchat_senders_ratio} further shows the loyal viewers who show up more often in livechats and members are more likely to send out Superchats. Across all VTubers, the median ratio of Superchat senders is only 5\%. Yet the same ratio is 14\% and 22\% for loyal viewers and members respectively. 

\begin{figure*}
    \centering
    \begin{subfigure}{\textwidth}
        \centering
        \includegraphics[width=0.7\linewidth]{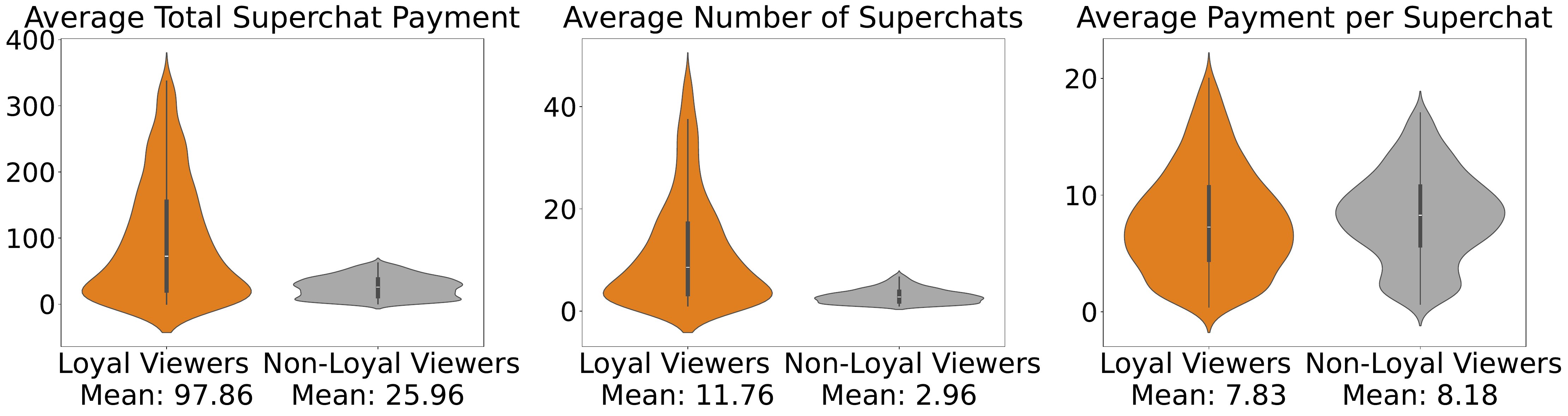}
        \caption{Statistics of Loyal Viewers' Superchat Income Contributions. Wilcoxon signed-rank tests revealed significant differences between loyal and non-loyal viewers regarding average total Superchat payment ($W = 30\ 874$, $p = 9.55\times 10^{-159}$), average number of Superchats ($W = 13\ 448.5$, $p = 7.56\times 10^{-174}$), and average payment per Superchat ($W = 272\ 609$, $p = 1.18\times 10^{-8}$).}
        \label{superchat_concentration_violin_plots_loyal_users}
    \end{subfigure}
    \begin{subfigure}{\textwidth}
        \centering
        \includegraphics[width=0.7\linewidth]{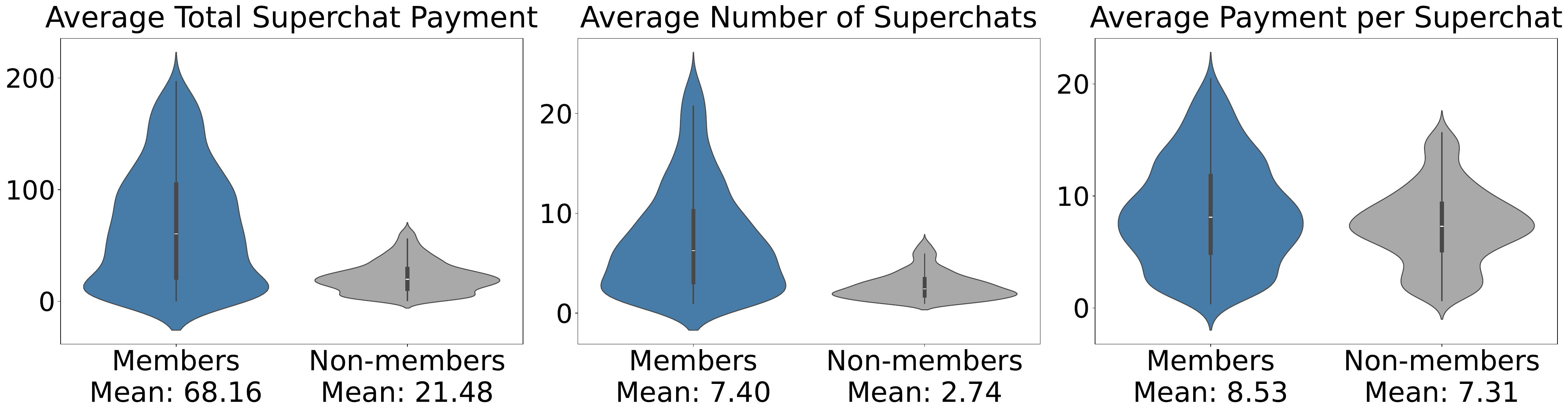}
        \caption{Statistics of Members' Superchat Income Contributions. Wilcoxon signed-rank tests revealed significant differences between members and non-members regarding average total Superchat payment ($W = 30\ 955$, $p = 1.59\times 10^{-127}$), average number of Superchats ($W = 23\ 010.5$, $p = 1.51\times 10^{-133}$), and average payment per Superchat ($W = 137\ 324$, $p = 4.24\times 10^{-35}$).}
        \label{superchat_concentration_violin_plots_members}
    \end{subfigure}
    \caption{Detailed Statistics of Superchat Income Contributions} 
    \label{fig:violin_plots_superchats_concentration}
\end{figure*}

We take a closer look at this disparity in Figure~\ref{fig:violin_plots_superchats_concentration}. On average, loyal viewers and members send more money through Superchats than non-loyal viewers and non-members (left), despite the former being outnumbered by the latter.  However, the USD-equivalent value of single Superchats sent by both loyal/non-loyal viewers and members/non-members was pretty close (right). The monetary gap is primarily caused by the fact that loyal viewers and members sent out Superchats far more frequently than other viewers.

\begin{figure*}
\centering
\begin{subfigure}{.4\textwidth}
  \centering
  \includegraphics[width=1\linewidth]{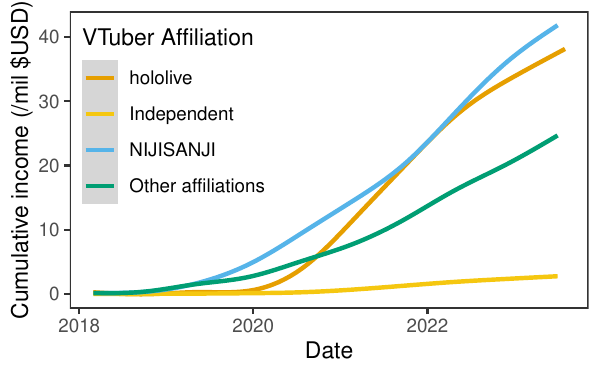}
  \caption{Cumulative Income of All VTubers by Affiliation}
  \label{fig:cumulative_income_by_affil}
\end{subfigure}%
\begin{subfigure}{.4\textwidth}
  \centering
  \includegraphics[width=1\linewidth]{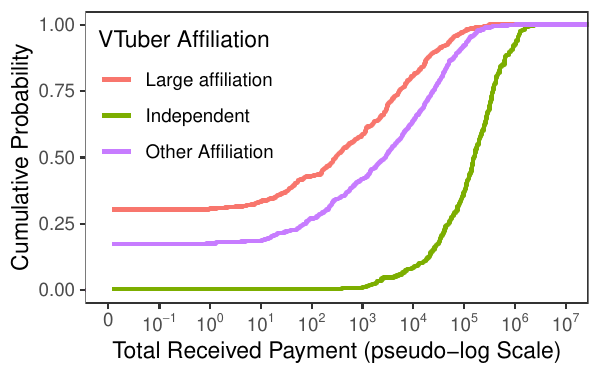}
  \caption{CDF of Income by Affiliation}
  \label{fig:cdf_affil}
\end{subfigure}
\caption{Cumulative Income and CDF of Income by Affiliation. The largest agencies (Hololive and NIJISANJI), dominate market share in terms of revenue.}
\label{fig:test}
\end{figure*}

\begin{figure*}
\centering
\begin{subfigure}{.33\textwidth}
  \centering
  \includegraphics[width=1\linewidth]{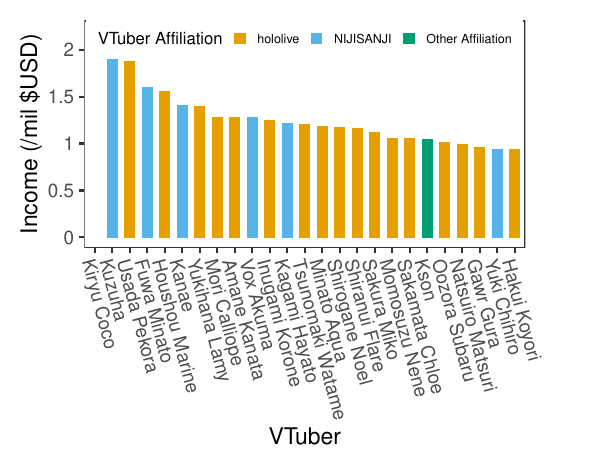}
  \caption{Top VTubers by Income}
  \label{fig:top_vtuber_bar}
\end{subfigure}%
\begin{subfigure}{.33\textwidth}
  \centering
  \includegraphics[width=1\linewidth]{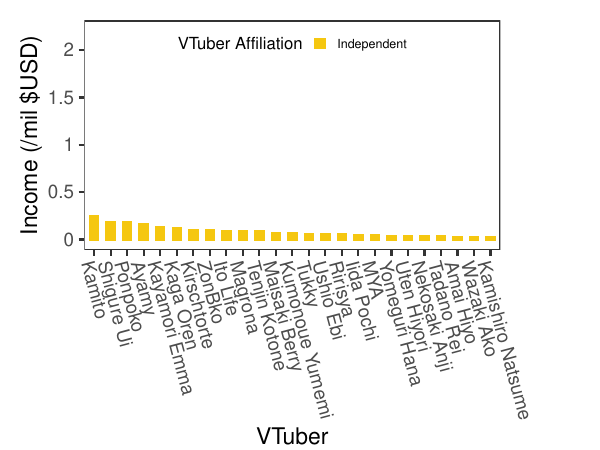}
  \caption{Top Independent Vtubers by Income}
  \label{fig:top_indie_vtuber_bar}
\end{subfigure}
\begin{subfigure}{.33\textwidth}
\centering
\includegraphics[width=1\linewidth]{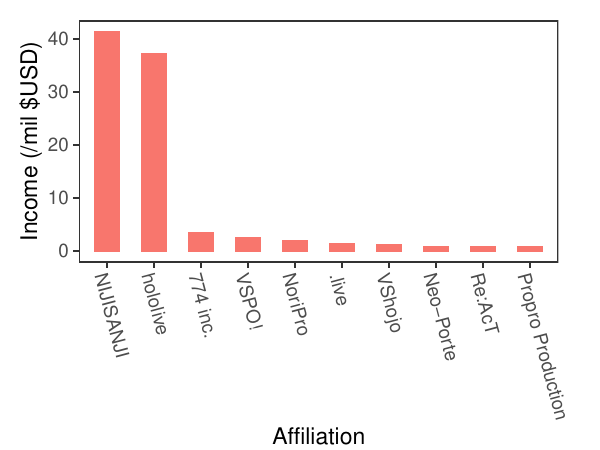}
\caption{Total Income of VTuber agencies}
\label{fig:top_affils}
\end{subfigure}
\caption{Incomes of Top VTubers and Top independent VTubers. The market is disproportionately controlled by the largest agencies.}
\label{fig:top_vtuber_incomes}
\end{figure*}

\section{RQ3: What is the influence of agencies in monetization?}
In this section, to address \textbf{RQ3}, we dive deeper into the role of agencies in the VTuber monetization landscape. 

\subsection{Affiliated VTubers hold a strong advantage over independent VTubers}
The top-earning VTubers are also dominated by corporate-managed VTubers, especially from two major VTuber agencies, Hololive and Nijisanji (see Figure~\ref{fig:gini}). Hololive dominates the leaderboard with more high-earning VTubers than Nijisanji and other affiliations (Figure~\ref{fig:top_vtuber_incomes}), though it manages fewer VTubers than Nijisanji (Table~\ref{tab:summary}). In contrast, the top independent VTubers earned far less than the top corporate VTubers (Figure~\ref{fig:top_vtuber_incomes}). For example, the top independent VTuber, \textit{Kamito}, only earned one eighth of the total income of the top VTuber, \textit{Kiryu Coco}, and was not even among the top 100 VTubers by income. 

We visualize the income disparity between affiliated and independent VTubers in the cumulative distribution function (CDF) of VTuber incomes in Figure \ref{fig:cumulative_income_by_affil}. Across the whole VTuber spectrum, independent VTubers earned less than affiliated VTubers, and a higher percentage of independent VTubers hadn't received any Superchat payment in their active period at all. 

\subsection{The VTuber market is dominated by two VTuber agencies}

Fig~\ref{fig:cumulative_income_by_affil} shows the growth of incomes by affiliations. Even among the VTuber agencies, the inequality is substantial.
\textit{Nijisanji Project} was the first major VTuber agency to reap significant profits from VTubing since around 2019. 
\textit{Hololive}, and most small agencies soon followed suit and began making large margins of profit in 2020. Since then, the overall growth has remained strong, catalyzed by the COVID-19 Pandemic \cite{polygonVtuberTakeover,yakura2021no,zhao2022live}. Notably, Nijisanji and Hololive, respectively, earn more than all other agencies combined in terms of Superchats, suggesting the corporate monopoly of the VTuber market. 

In Figure~\ref{fig:top_affils}, Hololive and Nijisanji have far outperformed the other agencies with more than 10 times the profits of the next largest agency. 
Even among affiliated VTubers, the VTuber agencies also differ widely in nature. Some are large corporate agencies specializing exclusively in managing VTubers like Hololive and Nijisanji with strict talent selection pipelines, professional content creation, and marketing strategies \cite{liudmila2020designing,zhao2022live}, while some are formed by a few independent VTubers loosely joining together to co-hire personnel or co-livestream. 

\subsection{Agencies took the lead in international expansion} 

\begin{figure*}
\centering
\begin{subfigure}{.33\textwidth}
  \centering
  \includegraphics[width=0.9\linewidth]{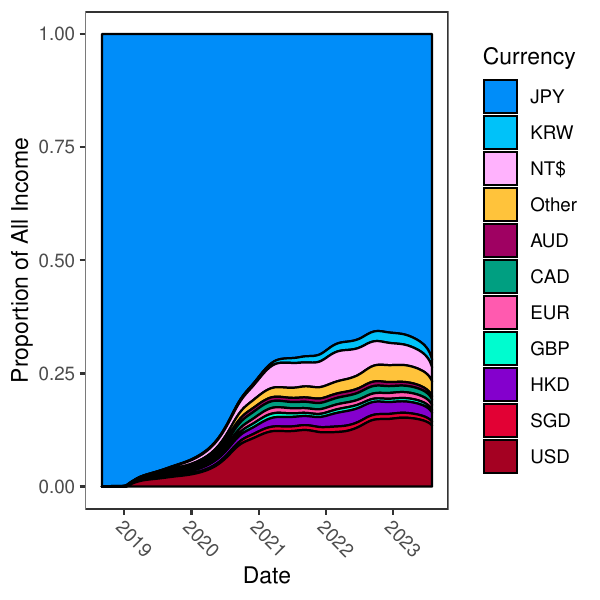}
  \caption{Temporal Distribution of Payments}
  \label{fig:temporal_payment}
\end{subfigure}%
\begin{subfigure}{.33\textwidth}
  \centering
  \includegraphics[width=0.9\linewidth]{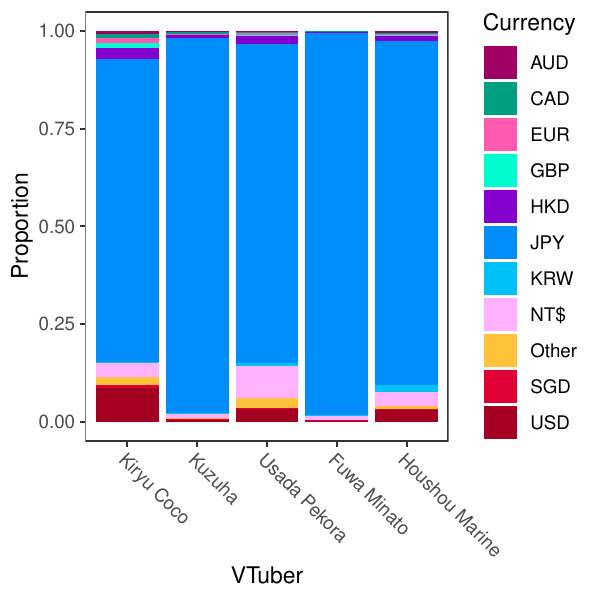}
  \caption{Currency Distribution of Top VTuber's}
  \label{fig:top_vtubers_sankey}
\end{subfigure}
\caption{Distribution of Currencies. VTubers' incomes consist of diverse international currencies.}
\label{fig:currency_composition}
\end{figure*}

\begin{figure*}
    \centering
    \includegraphics[width=0.8\linewidth]{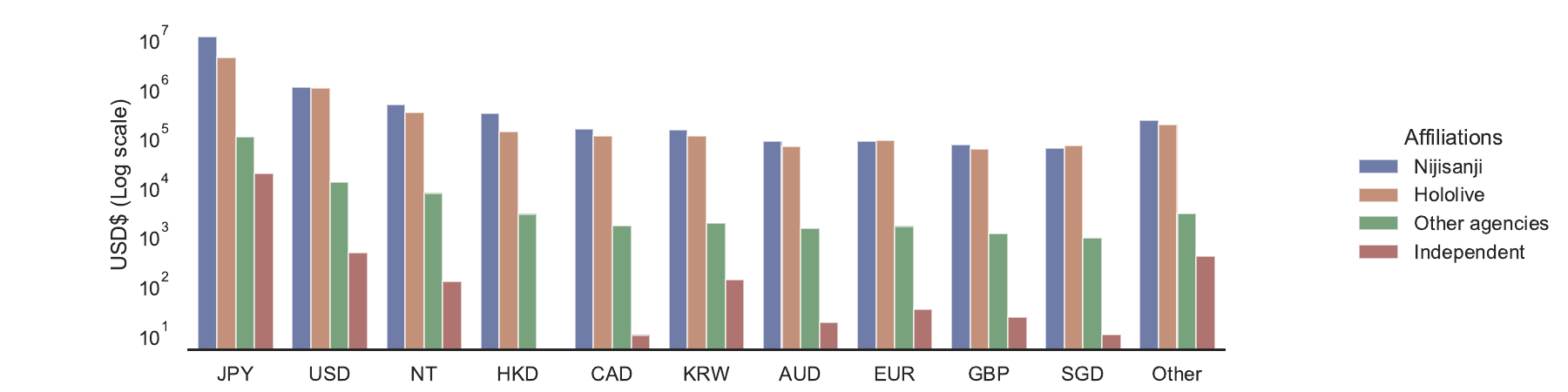}
    \caption{The USD equivalent values of different currencies by affiliation types. The results show that Hololive and Nijisanji consistently ranked as the top two across all currencies.}
    \label{fig:currency_grouped_by_agency}
\end{figure*}

Longitudinally, the income composition of currencies reflects the increasing International reception of VTubers. In Figure~\ref{fig:currency_composition}, the international expansion began in 2019 and has grown tremendously during the COVID-19 Lockdown \cite{zhao2022live,leith2022twitch}. Since then, while the primary profits of the VTuber industry are still from Japan, about 30\% of the profits were generated in other currencies. If these data were broken down by affiliations, we can observe in Figure~\ref{fig:currency_grouped_by_agency} that Hololive and Nijisanji ranked as the top two in almost all currencies, suggesting their leading positions in the international expansion.  

The growth of VTubers was primarily in East Asia, from Japan, Taiwan, Hong Kong, and South Korea, where Japanese ACG culture commands a large influence. North American countries (US and Canada) are the second largest regional income sources for VTubers, followed by European countries. Language similarity and cultural affinity play an important role in international expansion.

\begin{figure}
    \centering
    \includegraphics[width=\linewidth]{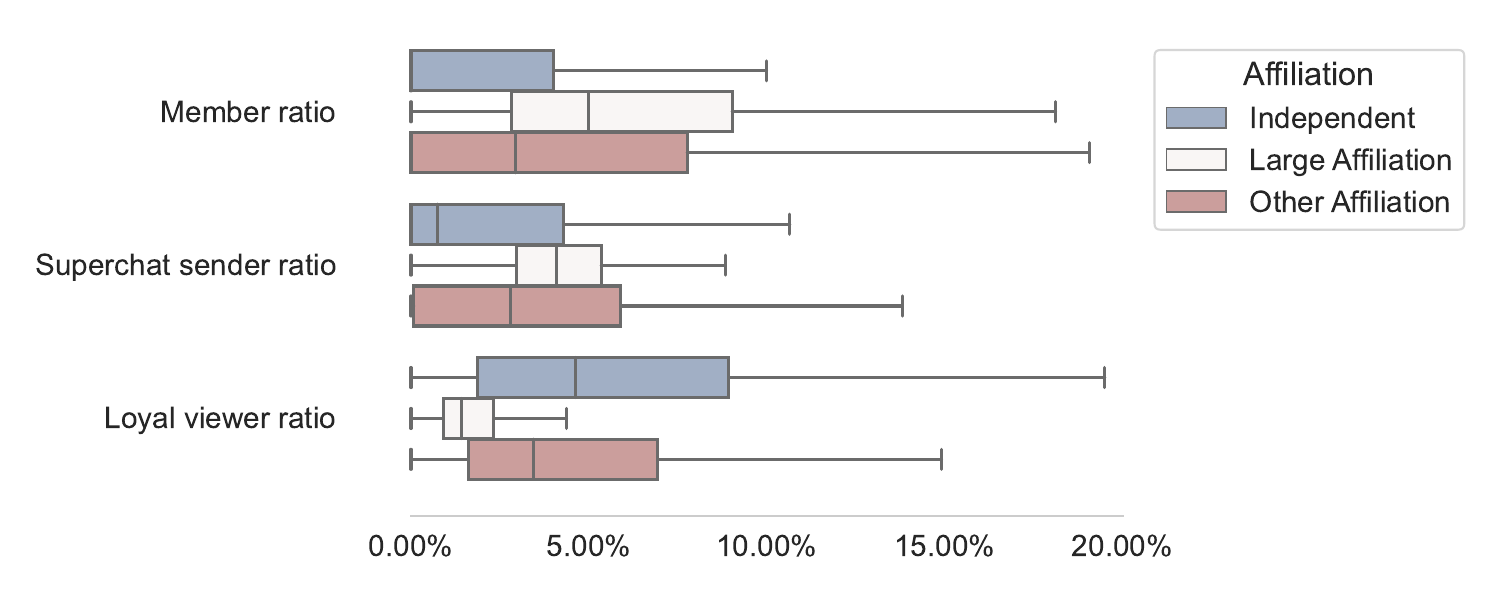}
    \caption{Ratio of different viewer category by affiliation types. Large agency VTubers had higher percentages of members and Superchat senders but lower percentages of loyal viewers compared to independent and other agency VTubers. Independent VTubers had higher percentages of loyal viewers but lower percentages of members and Superchat senders.
}
    \label{fig:viewer_ratio_grouped_by_agency}
\end{figure}


\subsection{VTubers from the same agency tend to attract similar loyal viewers}
Figure~\ref{fig:viewer_ratio_grouped_by_agency} shows the ratio of committed viewers by different affiliation types. We conducted Kruskal-Wallis H-tests to compare the differences within and across affiliation types. VTubers from large agencies attracted a higher percentage of members and Superchat senders than loyal viewers ($H=256.99$, $p=1.56e^{-56}$). In contrast, independent VTubers attract a higher percentage of loyal viewers than members and Superchat senders ($H=268.54$, $p=4.86e^{-59}$). VTubers from other agencies also have a higher ratio of loyal viewers than members and Superchat senders ($H=47.54$, $p=4.74e^{-11}$), though the difference is less pronounced. For each ratio, large agency VTubers had a lower percentage of loyal viewers than independent and other agency VTubers ($H=147.61$,	$p=8.80e^{-33}$), but had a higher ratio of members ($H=184.59$, $p=8.22e^{-41}$) and Superchat senders ($H=119.97$, $p=8.85e^{-27}$) than other VTubers. 

\begin{figure*}
    \centering
    \includegraphics[width=0.9\linewidth]{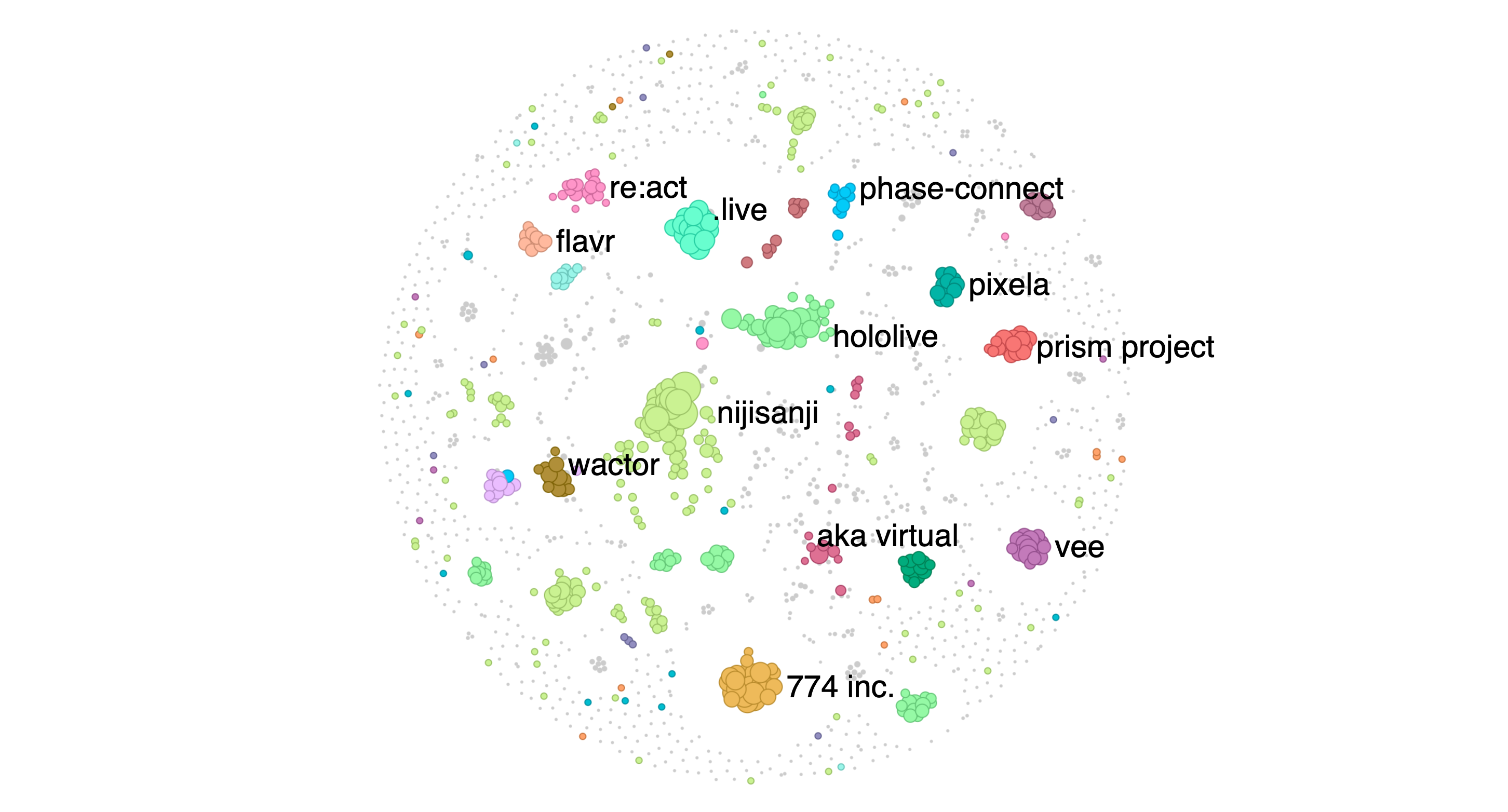}
    \caption{VTuber network constructed through shared loyal viewers. Each node represents a VTuber and node size is proportional to node degrees. Nodes connected through more edges are closer in distance. VTuber agencies are represented by colors. The network indicates that VTubers affiliated with the same agency tend to share a significant number of loyal viewers.}
    \label{fig:vtuber_graph_by_loyal_users}
\end{figure*}

Loyal viewers form the backbone of a fanbase, as they engage with VTubers more frequently and for a long time. The network shown in Figure~\ref{fig:vtuber_graph_by_loyal_users} illustrates the connections between VTubers through shared loyal users. This implies that loyal viewers of an affiliated VTuber are also more likely to be loyal to other VTubers from the same agencies. Nijisanji and Hololive also have multiple VTuber clusters, which are grouped by their local branches in Japanese, English, and Indonesian, or by gender of the VTubers.  Small agencies and independent VTubers tend to be less connected in the graph, suggesting that loyal viewers to these VTubers are more likely to be only loyal to them alone. Smaller VTubers were successful in building a fanbase of their own, yet agencies are keen on curating a fanbase for their agency brands rather than individual VTubers.

\section{RQ4: What factors affect the survival of VTubers?}
In this section, we examine the career lifespan of individual VTubers to answer \textbf{RQ4}. 

\subsection{Most VTubers fail within three years}
\begin{figure}
    \centering
    \includegraphics[width=\linewidth]{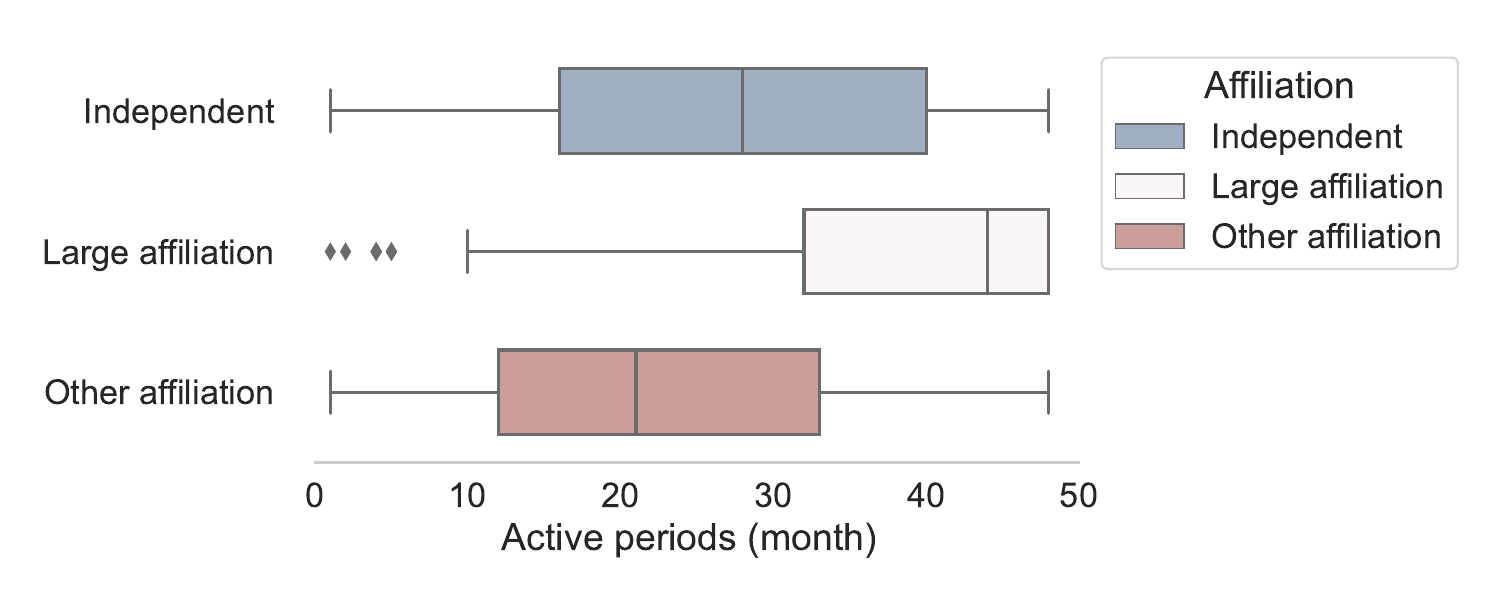}
    \caption{Distribution of VTubers' lifespan by affiliations. VTubers from the largest agencies tend to have the longest lifespan, followed by independent VTubers, and then those from smaller agencies.}
    \label{fig:lifespan_histograms}
\end{figure}

Figure~\ref{fig:lifespan_histograms} illustrates the distribution of VTuber lifespan, or the time difference between the first and the last streaming session in months. The median lifespan of all VTubers is 24 months. Further breaking down by affiliations indicates that the median active period is 44 months for large agencies (Hololive \& Nijisanji), 21 months for other agencies, and 28 months for independent VTubers. Pairwise Wilcoxon signed-rank test with False Discovery Rate correction \cite{benjamini1995controlling} shows that the differences are significant between each pair of affiliation type (Large vs. Other: $p=1.3e^{-28}$; Large vs. Independent: $p=2.07e^{-15}$; Independent vs. Other: $p=2.03e^{-7}$). 

While many VTubers will probably remain active after our cutoff time of July 2023, a high percentage of them have already stopped streaming long before the cutoff date. We find that 1,057 (over 50\%) have stopped streaming completely within 3 years of their first stream. Of those who fail, the median lifespan is about 19 months. Motivated by this observation, we continue to investigate what are the factors associated with VTubers' failure. 

\begin{table*}[]
\small
\begin{tabular}{lcccc}
\toprule
Feature & $\beta$ & exp($\beta$) & p-value & Significance \\ \midrule
Affiliation (Independent) & $0.259$ & $1.296$ & $0.302$ & \\
Affiliation (Non-large/other) & $0.683$ & $1.980$ & $0.003$ & ** \\
VTuber Gender (male) & $0.026$ & $1.026$ & $0.741$ & \\
VTuber Gender (other) & $0.139$ & $1.149$ & $0.277$ & \\
Total Stream Duration & $-0.026$ & $0.974$ & $0.207$ & \\
Total Superchat Income & $-0.136$ & $0.872$ & $0.111$ & \\
Total Number of Members & $0.753$ & $2.124$ & $1.6 x 10^{-8}$ & *** \\
Total Viewers & $-0.017$ & $0.983$ & $0.001$ & **\\
Affiliation (Independent) * time & $-0.019$ & $0.981$ & $0.027$ & * \\
Affiliation (Non-large/other) * time & $-0.019$ & $0.981$ & $0.015$ & * \\
Total Superchat Income * time & $-0.007$ & $0.993$ & $0.010$ & * \\ 
Total Number of Members * time & $-0.024$ & $0.976$ & $0.001$ & *** \\
Total Viewers * time & $0.0001$ & $1.000$ & $0.530$ & \\
\bottomrule
\end{tabular}
\caption{Cox Regression Model Summary.}
\label{tab:cox_model}
\end{table*}


\subsection{VTubers' failure is associated with affiliation status and channel performance}
Table \ref{tab:cox_model} summarizes the results of the Cox model regression fitted to VTuber survival data. We find that the role of affiliation, income, membership, and unique viewership are all statistically significant in predicting the hazard ratio for VTubers, either as main effects or as interaction terms with time. However, gender and streaming duration in the past month were not statistically significant predictors of survival hazard. We find that the mean concordance index across all folds is 0.680 ($\sigma = 0.023$). This is higher than 0.5 which is the expected concordance index for a random guessing baseline. This suggests that our features have predictive power over VTuber failure rate.

With respect to user engagement features related to a VTuber's streams, for each unit increase in income, and each unit increase in unique viewers, a VTuber's risk of failure decreases slightly. Interestingly, membership count appears to significantly increase the initial base risk of a VTuber's failure. However, membership and Superchat income have notable interactions with time which both show a decreased risk of failure that is magnified the latter into the VTuber's lifespan. This suggests that long-term VTuber survival is dependent on securing consistent Superchat income and viewership.

Affiliation plays a significant role in the survival of VTubers. The risk of failure for independent and other non-large affiliations is at almost 1.3 times and 2 times the baseline risk than those from large affiliations. Notably, the interaction between VTuber affiliation and time is statistically significant for all levels. For both independent and non-large affiliated VTubers, the risk of failure decreases over time, despite the high initial risk.

\section{Discussions}
Here we present an in-depth discussion of our findings. 

\subsection{Corporate monopoly of the VTuber market}
Our findings reveal the starting economic inequality between VTubers, similar to streamers in Twitch \cite{houssard2023monetization} and Chinese platforms \cite{liao2023cyber}, as the concentration of wealth emerges naturally in economic activities \cite{newman2005power,gabaix2009power}. 
Yet compared to prior studies on the inequality of livestreaming \cite{houssard2023monetization,liao2023cyber}, we further show that, while wealth is concentrated in a few top VTubers, most, if not all, of these top VTubers are also controlled by two agencies, Nijisanji and Hololive. These two agencies reap the most views and Superchats on YouTube, more than the total of all other VTubers combined in our dataset. The idol manufacturing process by the largest VTuber agencies was successful. Large agencies can afford more aesthetically pleasing avatars, better motion and face capture technologies, marketing strategies, content creation plans, and professional training \cite{liudmila2020designing,zhao2022live}, making them easily stand out from VTubers.  

Many MCNs optimize content creation through influence matrices \cite{liang2024manufacturing}, strategies also adopted by many VTuber agencies. Our analysis of VTuber networks reveals that many VTubers from the same agencies share a similar subset of loyal viewers. Such outcomes are more likely to emerge from the coordinated operations directed by agencies such as collaboration livestreaming rather than only driven by platform algorithms. This top-down manufacturing process also allows agencies to design niche content to target certain audience groups. For example, Hololive has regional branches catering specifically to English, Japanese, and Indonesian audiences, and it also has a special branch, HOLOSTARS featuring male VTubers.

Yet the monopoly of a few large agencies can also lead to some problems. Research has found that profit-driven live-streaming guilds in China \cite{zhang2019virtual} required their streamers to prioritize content that was deemed most profitable, limiting the content diversity and the creativity freedom of streamers. The profit-driven nature of agencies can potentially cause streamers to treat fans as assets and minimize the reciprocal relationship \cite{zhang2019virtual,wang2019love}. They might also push independent VTubers and VTubers from small agencies out of the market.

\subsection{Precarity of the VTuber profession}
For most people, being a VTuber is far from being profitable and stable. In the platform `gig economy` \cite{morgan2018creativity,vallas2020platforms}, most streamers face economic instabilities and can be subject to various challenges posed by the platform, including demonetization and algorithmic biases \cite{kingsley2022give}.  
Agencies alter the power dynamics between streamers and the platform, making streamers easier to prosper and more resistant to the exploitation of the platform. 

However, affiliation with an agency does not necessarily benefit the VTuber career. Our finding suggests that independent VTubers were active for a slightly longer period than VTubers affiliated with small agencies other than Hololive and Nijisanji. Agencies are set up primarily for profit, yet a potion of independent VTubers are also motivated by intrinsic interests \cite{wang2019love}. Intrinsic interests can motivate some Independent VTubers to persist longer even if the monetary prospect is dim. Yet the pressure to monetize in small agencies might be more likely to cause burn-out and the subsequent dropout for some affiliated VTubers.  

While VTuber agencies shield VTubers from platform risks, they could also exploit VTubers. As the agencies own the anime avatar, the YouTube channel and the content, voice actors can be reduced to replaceable assets of the company. Contractual VTubers may even be subject to unfair labor practices, including delayed payment, harassment, and toxic work environments \cite{Hashimoto_2023}. Nijisanji also once publicly announced the graduation of a VTuber without even informing her \cite{polygonVTuberSelen}.
Secondly, the corporate agency usually takes a large share of the income to cover other costs, so what is left to VTubers could be a small percentage of what is actually earned \cite{liang2024manufacturing,liu2023zhibo}.

\subsection{Cultivating parasocial relationship from viewers}
In our results, viewers who were committed either financially (members) or behaviorally (loyal viewers) were more likely to be Superchat senders. For small VTubers, their streaming sessions tend to have less audience, making it possible for the VTuber to interact more frequently with viewers with high self-disclosure \cite{tan2025can} and for the viewers to perceive the VTuber as more authentic. It is far easier to cultivate the `perceived interconnectedness` \cite{abidin2015communicative} between viewers and the streamer. 

Corporate streamers often minimized the creation of reciprocal relationships but were more oriented towards maximizing the virtual gifts \cite{zhang2019virtual}. The business model of large agencies makes it harder to cultivate such relationships, even though their affiliated VTubers were more successful in attracting more members and Superchat senders. Each livestreamming session was planned and monitored by agency staff and VTubers were trained to behave in ways that could attract more potential monetization. Compared to independent VTubers, corporate VTubers tend to have less self-expression and are more distant to the audience \cite{tan2025can}. 

For agencies, the strategy is to turn loyal viewers into loyal customers of their agency brand through coordinated marketing. Many agencies marketed their VTubers as classmates or families and coordinated many collaborative streaming events where several VTubers interact closely. So viewers are more likely to extend their attachment from one VTubers to other VTubers from the same agency.

\subsection{Theoretical implications} 
Our study advances the general understanding of the creator economy, especially the monetization outcome of corporate players in institutionalizing content production. In the research on the platform economy, the major players are generally conceived as the platform, creators, users, and advertisers \cite{bhargava2022creator,peres2024creator,bleier2024role}, but the critical role of MCNs has not been sufficiently recognized. In many analyses, the assumption that individual channels or streamers are isolated members or isolated units of analysis is still widely held \cite{houssard2023monetization,rieder2023making,tan2025can}. Yet our analyses show that, agencies have dominated, if not monopolized, the VTuber market, exerting an unnegligible impact on the monetization landscape. 
Our study reveals the depth and breadth of corporate power from a quantitative perspective, challenging the `gig economy' assumption of the content creator economy. Future theoretical and empirical research should take MCNs into consideration. We call for more research to investigate how corporate forces impact and reshape the creator economy.

\subsection{Design implications}
\subsubsection{Better governance of agencies}
Large monopoly agencies, with their strong capability to attract viewers and to monetize, might also be able to challenge or strengthen the power of the platform \cite{zhang2024contesting} or gate-keep the VTuber market \cite{liang2024manufacturing}. From our analysis, agencies are extremely successful at attracting high ratios of paid viewers and large spikes of payment events. Yet it is not uncommon for agencies or MCNs to gain unfair advantage through gaming the algorithm, faking Superchat transactions, and encouraging impulse buying \cite{liu2023zhibo}. It is important to consider these agencies in platform governance to create a fair environment for content creators.

\subsubsection{Supporting independent VTubers} 
While the majority of small VTubers do not contribute to most views and revenues, it is still important to provide support for them on the platform. They produce most content on the platform, assume the risk of innovation, and put competitive pressure on the top VTubers and agencies \cite{rieder2023making}. Support efforts can include more accessible data analytics for tracking channel performance (like our analysis here), developing low-cost technologies for avatars and motion capture, and more equitable platform algorithms or profit-sharing programs to give more economic opportunities to independent VTubers.

\subsubsection{Enhancing content diversity} 
Due to the limitation of the anime avatar, the activities a VTuber can perform are much more limited than a real-person streamer. Our analysis showed that VTubers overwhelmingly rely on gaming, chatting, and singing, a standard production pipeline first codified by Nijisanji \cite{tan2025can}. 
It is necessary for VTubers to explore more diverse content creation strategies that differ and improve from the common VTuber streaming. For example, VTuber \textit{kson ONAIR} alternates between a virtual avatar and a real person in streaming. Such explorations in new content can cultivate a more diverse audience outside the Otaku community, and promote a more sustainable content creation. 

\subsubsection{Enhancing engagement with multilingual technologies}
The international audience has been an important source of VTubers' monetization. We tried to access some top VTuber pages by setting our location to different regions and languages, only to find that little was done on the side of VTubers or the platform to enhance the accessibility of these international audiences other than the localized livechat interface.  Technologies like streaming speech recognition and machine translation are promising in partially bridging the language gap during livestreaming. The incorporation of multilingual language technologies can potentially attract more international viewers for VTubers.

\section{Limitations}
Our study is still limited in several ways. First, our results can still be biased, as samples cannot represent a complete history of VTubers and their viewers on YouTube. Some VTubers chose not to make public the past streamed sessions due to privacy, copyright concerns, and contractual requirements. So we had to exclude them in our analysis. There are also other popular VTuber platforms, like Twitch, Bilibili (Chinese VTubers), and NicoNico (Japanese VTubers). Secondly, we only included viewers who left livechats, which can be just a small fraction of all viewers. Thirdly, our estimation of VTuber incomes is only an estimation based on Superchats, which might only account for a part of the actual incomes of VTubers. The same USD amount also might not represent the same buying power in different regions.
Finally, like other data mining studies, our statistical analysis only reveals correlation rather than causal relationships. Future research is needed to address these limitations to better understand VTubers or streamers in general. 

\section{Conclusions}
In this study, we analyzed over a million hours of publicly available streaming records of over 1900 VTubers on YouTube to understand their monetization landscape. Our analysis reveals stark inequality and instability of incomes for the majority of VTubers behind the booming industry, as only a few top VTubers and two VTuber agencies are dominating the whole market. We also characterize the viewers and the role of agencies. Furthermore, survival analysis suggests that VTubers' failure can be predictable based on economic factors and affiliation status. For the HCI community and relevant stakeholders, these results shed light on the overall functioning of the VTuber ecosystem and provide implications informing the design of sustainable support systems for VTubers. Our research also calls for more research on the impact of corporate forces on the creator economy. 

\begin{acks}
We thank the ACs and the anonymous reviewers for their detailed comments, which helped improve this article considerably. 
We also thank UBC Advanced Research Computing and Digital Alliance of Canada for their computing support. BD was supported by the Canada Graduate Scholarship (Master's Program). This research was funded by the Hampton New Faculty Grant, the CFI-JELF Fund, and the NSERC Discovery Grant awarded to the last author JZ.
\end{acks}

\bibliographystyle{ACM-Reference-Format}



\end{document}